\documentclass[a4paper,UKenglish,cleveref, autoref, thm-restate]{lipics-v2021}

\hideLIPIcs  
\nolinenumbers

\title{Convergence and Running Time of Time-dependent Ant Colony Algorithms}

\author{Bodo Manthey}{Faculty of Electrical Engineering, Mathematics, and Computer Science, University of Twente, The Netherlands \and \url{https://people.utwente.nl/b.manthey} }{b.manthey@utwente.nl}{https://orcid.org/0000-0001-6278-5059}{}
\author{Jesse van Rhijn}{Faculty of Electrical Engineering, Mathematics, and Computer Science, University of Twente, The Netherlands \and \url{https://people.utwente.nl/j.vanrhijn}}{j.vanrhijn@utwente.nl}{https://orcid.org/0000-0002-3416-7672}{Supported by NWO grant OCENW.KLEIN.176.}
\author{Ashkan Safari}{Department of Quantitative Economics, School of Business and Economics, Maastricht University, The Netherlands \and \url{https://ashkansafari.com/}} {a.safari@maastrichtuniversity.nl}{https://orcid.org/0000-0002-0471-1884}{Supported by NWO grant OCENW.KLEIN.176.}
\author{Tjark Vredeveld}{Department of Quantitative Economics, School of Business and Economics, Maastricht University, The Netherlands \and \url{https://www.maastrichtuniversity.nl/nl/t-vredeveld} }{t.vredeveld@maastrichtuniversity.nl}{https://orcid.org/0000-0002-6357-8622}{}

\authorrunning{B.\ Manthey, J.\ van Rhijn, A.\ Safari and T.\ Vredeveld} 

\Copyright{Jane Open Access and Joan R. Public} 

\ccsdesc[500]{Theory of computation~Random search heuristics}
\ccsdesc[500]{Theory of computation~Random walks and Markov chains}
\ccsdesc[500]{Theory of computation~Shortest paths}

\keywords{metaheuristics, ant colony optimization, shortest path problem} 

\category{} 

\relatedversion{} 

\EventEditors{John Q. Open and Joan R. Access}
\EventNoEds{2}
\EventLongTitle{42nd Conference on Very Important Topics (CVIT 2016)}
\EventShortTitle{CVIT 2016}
\EventAcronym{CVIT}
\EventYear{2016}
\EventDate{December 24--27, 2016}
\EventLocation{Little Whinging, United Kingdom}
\EventLogo{}
\SeriesVolume{42}
\ArticleNo{23}

\newcommand{\probP}{\ensuremath{\mathbb{P}}}
\newcommand*\dd{\mathop{}\!\mathrm{d}}

\newcommand{\exP}{\mathbb E}
\newcommand\whp{\emph{whp}}
\newcommand{\A}{\mathcal{A}}
\DeclareMathOperator{\li}{Li}

\usepackage{mathtools}
\usepackage{tikz}
\usepackage{tkz-graph}
\usetikzlibrary{calc, math, decorations.markings}
\tikzstyle{vertex}=[circle, fill, inner sep=0pt, minimum size=6pt]
\newcommand{\vertex}{\node[vertex]}

\DeclarePairedDelimiter\ceil{\lceil}{\rceil}
\DeclarePairedDelimiter\floor{\lfloor}{\rfloor}

\usepackage{todonotes}

\begin{document}

\maketitle

\begin{abstract}

Ant Colony Optimization (ACO) is a well-known method inspired by the foraging behavior of ants and is extensively used to solve combinatorial optimization problems.
In this paper, we first consider a general framework based on the concept of a construction graph -- a graph associated with an instance of the optimization problem under study, where feasible solutions are represented by walks.
We analyze the running time of this ACO variant, known as the Graph-based Ant System with time-dependent evaporation rate (GBAS/tdev), and prove that the algorithm’s solution converges to the optimal solution of the problem with probability 1 for a slightly stronger evaporation rate function than was previously known.
We then consider two time-dependent adaptations of Attiratanasunthron and Fakcharoenphol's $n$-ANT algorithm: $n$-ANT with time-dependent evaporation rate ($n$-ANT/tdev) and $n$-ANT with time-dependent lower pheromone bound ($n$-ANT/tdlb). We analyze both variants on the single destination shortest path problem (SDSP). Our results show that $n$-ANT/tdev has a super-polynomial time lower bound on the SDSP. In contrast, we show that $n$-ANT/tdlb achieves a polynomial time upper bound on this problem.

\end{abstract}

\section{Introduction}

Ant Colony Optimization (ACO) is a nature-inspired metaheuristic first proposed by
Dorigo, Maniezzo and Colorni~\cite{dorigo1991ant}. The heuristic attempts to model the
capacity of ants to find efficient paths between their nest and food sources, effectively
solving shortest-path problems without intrinsic knowledge of the problem structure
\cite{gossSelforganizedShortcutsArgentine1989}.
This property - zero or little knowledge of the problem - is a strongly desired
property of general purpose optimization algorithms.
In nature, ants communicate by depositing pheromones on the ground while searching their environment for food sources. These pheromone trails attract other ants, who are inclined to follow the paths established by their predecessors. When foraging ants discover multiple routes between the nest and a food source, shorter paths become infused with pheromones more rapidly than longer ones because ants can traverse them more quickly. Consequently, as more ants choose the shorter path, the pheromone concentration along it increases, eventually leading almost all ants to follow this most efficient route.

Since its inception, different variants of ACO have been proposed and implemented,
to different levels of success, and have been applied to different problems such as Shortest Paths~\cite{Attiratanasunthron2008running, sudholt2012running}, Traveling Salesperson Problem~\cite{dorigo1999ant, Kötzing2012, stutzle1997max, zhou2009}, Minimum Weight Vertex Cover~\cite{shyu2004ant}, Minimum Cut~\cite{kotzing2010ant}, and Minimum Spanning Tree~\cite{neumann2010ant}. 

As is often the case with heuristics,
theoretical guarantees of its performance came later. The first convergence results
were obtained by Gutjahr~\cite{gutjahr2000graph}, analyzing a variant called the
\emph{Graph-based Ant System} (GBAS). 
In summary, GBAS sets up a graph $\mathcal{C}$ (called the \emph{construction graph}) given
an instance $\Pi$ of a combinatorial optimization problem.  One maps solutions
of $\Pi$ to walks in $\mathcal{C}$ starting and ending at fixed nodes $s$ and $t$.
A run of GBAS is divided into \emph{cycles}. In each cycle,
a number of ants (chosen by the practitioner)
are placed on a special starting node of $\mathcal{C}$ and perform a random walk.
When an ant finds a walk $w$ from $s$ to $t$, it computes the cost of the corresponding solution.
Then some amount of pheromone is placed on the arcs of the construction graph, in such a way
that good walks are reinforced in the next cycle. The specific rule according to which
the pheromone is placed differs per GBAS variant.
Simultaneously, some
arcs lose some pheromone at a set rate, the \emph{evaporation rate} $\rho \in (0, 1)$.

In a first paper \cite{gutjahr2000graph}, Gutjahr showed that GBAS finds the optimal solution of
any problem with a unique optimum with probability $1 - \varepsilon$, which he later generalized to problems
with non-unique optima~\cite{gutjahr2003generalized}. Here,
$\varepsilon \in (0, 1)$ is a constant fixed by the number of ants,
the size of the optimal solution, and the evaporation rate.
Also with probability $1 - \varepsilon$, the pheromone values on the arcs of $\mathcal{C}$ converge to
zero everywhere except on the arcs corresponding to the optimal walk $w^*$.
Crucially,
$\varepsilon$ does not depend on the number of cycles $m$ for which the algorithm is run,
and hence GBAS cannot be guaranteed to find the optimum even in the limit
$m \to \infty$. 
Contrastingly, in a simple random walk where each ant chooses the next node in its walk
uniformly at random, the time taken for any ant to find the optimal walk is finite with
probability 1.

Many variants of ACO have been described in the literature.
A brief overview is given by Dorigo and St\"utzle 
\cite[Chapter 2]{dorigo2004ant}.
These variants can be rather involved, providing functionality
for more complicated pheromone update rules and centralized ``daemon actions''.
Although this makes ACO a very flexible heuristic, it complicates the analysis.
We therefore restrict our attention to simpler variants for the purposes of this paper.

One variant that received decent attention is the \emph{Max-Min Ant System} (MMAS)
\cite{stutzle2004maxmin}.
This version of ACO mainly differs from GBAS in restricting the range of possible
pheromone values on any arc to $[\tau_{\min{}}, \tau_{\max{}}]$ and
in the pheromone update rule, where the best solution found thus far is
reinforced in every cycle. The parameters
$\tau_{\min}$ and $\tau_{\max}$ are chosen in advance, but may depend on the
problem instance. The running time of this variant has been analyzed
for classical shortest path problems
\cite{Attiratanasunthron2008running,sudholt2012running}, achieving polynomial
running time there, and on the Traveling Salesperson Problem \cite{Kötzing2012,zhou2009}, yielding high-quality
solutions in polynomial time.

Extending both GBAS and MMAS, Gutjahr \cite{gutjahr2002aco} proposed variants
for both of these algorithms. First, he proposed GBAS/tdev, for \emph{time-dependent
evaporation rate}. This algorithm replaced the constant evaporation rate $\rho$
by a decreasing function $\rho : \mathbb{N} \to (0, 1)$. For a suitable function, this results in
convergence to an optimal walk $w^*$ with probability 1. Second, he proposed GBAS/tdlb, for
\emph{time-dependent lower bound}. This algorithm is similar to
MMAS, but replaces $\tau_{\min}$ by a decreasing function $\tau_{\min}(k)$.
Again, an appropriate choice of this function yields convergence with
probability 1.
Gutjahr left as an explicit open problem to analyze the running time of
these algorithms on specific problems. Later, he
analyzed the running time of GBAS/tdlb on a variant of the OneMax problem, a simple pseudo-boolean
optimization function~\cite{GUTJAHR20082711}. We are not aware of running time results of this algorithm
on more natural optimization problems, nor of any such results on GBAS/tdev.

Meanwhile, Attiratanasunthron and Fakcharoenphol \cite{Attiratanasunthron2008running}
analyzed the running time of a variant of MMAS on the single-destination shortest path problem
(SDSP) on directed acyclic graphs (DAGs). In this work, they first identified a shortcoming
of the usual MMAS algorithm where each ant starts from the same source vertex: They construct a
graph for which MMAS takes exponential time to solve the SDSP, even though the problem can
be solved efficiently using, e.g., the Bellman-Ford algorithm.
To solve this problem, Attiratanasunthron and Fakcharoenphol propose the $n$-ANT algorithm, which is inspired by the 1-ANT~\cite{neumann2009runtime} and the AntNet algorithm~\cite{di1998antnet}. This variant
of MMAS places an ant on every node of the input DAG. Each ant independently searches for the
target node, and is only responsible for updating the pheromone value of the outgoing edges
of its source node. For a DAG with $n$ nodes and $m$ edges, they obtain a running time bound
of $O\left(\frac{1}{\rho}n^2 m \log n\right)$. Later, Sudholt and Thyssen~\cite{sudholt2012running}
modified this algorithm to work on more general classes of graphs,
and obtained an improved running time bound.

\subsubsection*{Our Contributions}

In this paper, we obtain a running time bound for GBAS/tdev, and additionally improve the
evaporation rate function required for convergence. We also analyze the running time of time-dependent
variants of $n$-ANT, showing both a lower bound (for a time-dependent evaporation rate)
and an upper bound (for a time-dependent lower pheromone bound). A more
detailed explanation of our results follows.

Inspired by Gutjahr's time-dependent algorithms, we first consider the running time
of GBAS/tdev on general optimization problems. Rather than only considering the limiting
behavior of the algorithm, we obtain a finite-time bound for the 
expected number of cycles
necessary for the pheromone on every edge not in an optimal walk to decrease to some small
threshold. The analysis uses similar ideas already used by Gutjahr; however, we
obtain a small improvement in the sequence of evaporation rates $\rho(k)$ needed to
guarantee convergence. 

For the following theorem, let $S$ denote the
number of ants used in a run of GBAS/tdev, let $n$ be the number of nodes in the
construction graph of GBAS/tdev for some optimization problem (see \Cref{sec: gbas specification}
for a definition), and let $L$ be an upper bound to the number of nodes
in an optimal walk.

\begin{restatable}{theorem}{gbasmainresult}\label{thm:gbas running time}
    Let $\rho(k) = \dfrac{\alpha}{k}$ be the value of the evaporation rate used by GBAS/tdev in cycle $k$, where
    $\alpha < \dfrac{1}{2L}$. Fix some $\varepsilon > 0$. Then there exists a constant
    $c>0$ such that, if $m^{1-2cL} \geq c\cdot \frac{n^{2L}}{|S|} \cdot \ln \frac{1}{\varepsilon}$,
    the probability that no ant traverses an optimal walk in GBAS/tdev
    in the first $m$ cycles is bounded from above by $\varepsilon$.
\end{restatable}

Next, we turn to Attiratanasunthron and Fakcharoenphol's $n$-ANT algorithm~\cite{Attiratanasunthron2008running}, and combine
it with Gutjahr's time-dependent evaporation rate (tdev) and time-dependent lower bound (tdlb) 
variants of GBAS to obtain
$n$-ANT/tdev and $n$-ANT/tdlb. We first show that $n$-ANT/tdev has
a super-polynomial running time on the SDSP.


\begin{restatable}{theorem}{nantrunningtimetdev}\label{thm: nant running time tdev}
    For all $n \geq 3$, there exists an instance of the SDSP on $n$
    vertices such that $n$-ANT/tdev takes $2^{\Omega(\sqrt n)}$ cycles until all ants
    have seen their shortest paths, provided that the evaporation rate function has
    the form $\rho(m) = \alpha_n/m^\beta$ for $\alpha_n < 1$ and $\beta > 0$, where $\alpha_n$ may be a non-increasing function of $n$ and $\beta$ is a constant.
\end{restatable}

This tells us that the `tdev' modification
may cause a large running time for some ACO variants. 

On the positive side, we show that for the same sequence of decreasing $\tau_{\min}$ values
proposed by Gutjahr~\cite{gutjahr2002aco}, $n$-ANT/tdlb has polynomial expected running time. Compared
to the original definition of the algorithm, this variant has the nice theoretical
property that, as the number $m$ of cycles goes to infinity, the pheromone values only
retain a non-zero value on edges which are part of a shortest path to the target node.
In the following, let $n$ be the number of nodes in an instance $\mathcal{G}$ of the SDSP,
and let $\tau_{\min}(m) = c_n / \log(m+1)$ (as recommended by Gutjahr~\cite{gutjahr2002aco}).

\sloppy\begin{restatable}{theorem}{nantrunningtime}\label{thm:totalCycles}
    Let $\mathcal{G} = (\mathcal{V}, \mathcal{A})$ be an instance of the SDSP
    on which we run $n$-ANT/tdlb with 
    $\tau_\text{min}(m) = c_n /\ln(m + 1)$. The expected number of cycles until all ants have found their shortest paths from their respective starting
    nodes in $\mathcal{G}$ is 
    %
    \[
        O\left(\dfrac{n^2 \cdot \tau_\text{max}}{c_n} \cdot \ln{\left(\dfrac{n \cdot \tau_\text{max}}{c_n}\right)} + \dfrac{n}{\rho} \ln(n\cdot\tau_\text{max})\right),
    \]
    for $\tau_{max} \geq 1$ and $0 < c_n \leq \frac{1}{n^2}$.
\end{restatable}

If we set $\tau_{\max} = 1$, and $c_n = \frac{1}{n^2}$, then the number
of cycles amounts to $O\left(n^4 \ln n +\dfrac{n}{\rho} \ln n\right)$.

The following corollary to \Cref{thm:totalCycles} shows that the
pheromone values converge to zero everywhere except on some solution to
the SDSP.

\begin{restatable}{corollary}{nantlimitinggraph}
\label{cor:limiting graph}
    Let $\mathcal{G} = (\mathcal{V}, \mathcal{A})$ be an instance of the SDSP. Let
    $\underline{\tau}(m) \in \mathbb{R}^{\mathcal{A}}$ be the vector of pheromone values in cycle
    $m$ of a run of $n$-ANT/tdlb. Let $\mathcal{G}_\delta(m)$ be the graph with
    the same node set as $\mathcal{G}$, but retaining only those edges
    $e \in \mathcal{A}$ such that
    $\tau_e(m) > \delta$. Then for each $0 < \delta < \tau_{\max}$,
    the limiting graph $\lim\limits_{m \to \infty} \mathcal{G}_\delta(m)$ has an edge set equal to an optimal
    solution of the SDSP with probability 1.
\end{restatable}

We note that the $n$-ANT variant defined by Sudholt and Thyssen~\cite{sudholt2012running} has a better
running time on the SDSP than the original version as proposed by Attiratanasunthron and Fakcharoenphol~\cite{Attiratanasunthron2008running}.
We chose not to analyze this version, as the analysis
is somewhat more involved, and our aim is only to show that the tdlb
method can yield a polynomial running time.



\section{Graph-based Ant System with Time-dependent Evaporation Rate}


Recall that GBAS with a constant evaporation rate cannot guarantee that the ants find an optimal solution even
in the limit where the running time approaches infinity. This is true typically for
ACO algorithms which do not impose a lower limit on the pheromone values on the arcs, including $n$-ANT (see
\Cref{sec:nant}).
Intuitively, this is because the arcs of an optimal walk may lose too much pheromone too quickly,
which decreases their probability of being chosen. This effect is self-reinforcing, because the
arcs which are not chosen in any cycle lose some pheromone.

It is desirable to have an ACO algorithm which guarantees convergence to an optimal walk with
probability 1. To achieve this goal, Gutjahr~\cite{gutjahr2002aco} took inspiration from \emph{simulated
annealing}~\cite{aartskorst2003, kirkpatrick1983},
another randomized metaheuristic, which does guarantee convergence to an optimum. Stronger still,
simulated annealing is described by a Markov chain with a limiting distribution that places positive
probability \emph{only} on optimal solutions. Key to the approach is to vary a parameter,
the \emph{temperature}, during the execution of the algorithm. By decreasing the temperature to zero
sufficiently slowly, convergence is achieved with probability 1.

Applying this approach to ACO, Gutjahr defined two new variants: one with a time-dependent evaporation
rate (called GBAS/tdev) and one with a time-dependent lower pheromone bound (GBAS/tdlb)~\cite{gutjahr2002aco}.
Both approaches yield ACO variants that guarantee convergence to optimal solution, with the property
that in the limit, the pheromone values only retain positive values on the arcs of some optimal solution.

\subsection{Algorithm Specification}
\label{sec: gbas specification}

In this subsection, we formally describe GBAS/tdev, with a similar presentation to that
in Gutjahr's works~\cite{gutjahr2000graph, gutjahr2002aco, gutjahr2003generalized}. 
GBAS/tdev represents a feasible solution to an instance $\Pi$ of a combinatorial optimization problem as a walk in a directed graph $\mathcal{C} = (\mathcal{V}, \mathcal{A})$, referred to as the ``construction graph''.
In this graph, a single node is designated as the start node.
Let $\mathcal{W}$ be the set of directed walks in $\mathcal{C}$. Each walk $w \in \mathcal{W}$ starts at the start node of $\mathcal{C}$, and it contains each node of $\mathcal{C}$ at most once; this implies that the walks in $\mathcal{W}$ are finite.
We also have a function $\Phi$ that maps the set of walks in $\mathcal{C}$ onto a set of solutions to $\Pi$ including all the feasible solutions (note that we therefore also map some walks to \emph{infeasible} solutions).
Therefore, $(\mathcal{C}, \Phi)$ specifies an encoding of the solutions to $\Pi$ as ``walks'' in $\mathcal{C}$. 
The value of the objective function of a walk in $\mathcal{C}$ is equal to the value of the objective function of the corresponding solution of $\Pi$ if this solution is feasible; otherwise, it is set to infinity. We assume in this paper that the optimization problem under consideration is a minimization problem.

In GBAS/tdev, we have a set $S$ of \emph{ants} (also called \emph{agents} in the literature).
Each ant $j \in S$ performs a random walk in $\mathcal{C}$ according to some transition probabilities.
A \emph{cycle} is a time period in which all the ants perform a random walk in $\mathcal{C}$.
The transition probability of going from node $i$ to node $i'$ in $\mathcal{C}$ in cycle $m$ is
\begin{align*}
    p_{i,i'}(m) = \dfrac{\tau_{i, i'}(m)}{\sum_{i'': (i, i'') \in \mathcal{A}} \tau_{i, i''}(m)},
\end{align*}
where $\tau_{i, i'}(m)$ is the \emph{pheromone value} of the arc $(i, i')$ in cycle $m$.
At the beginning of cycle 1, we initialize all the pheromone values by $\frac{1}{|\mathcal{A}|}$; i.e., $\tau_{i, i'}(1) = \frac{1}{|\mathcal{A}|}$ for all $(i, i') \in \mathcal{A}$.
At the end of each cycle $m$, the pheromone value of an arc $(i, i')$ is updated as follows: 
\begin{align*}
    \tau_{i, i'}(m+1) = 
    \begin{cases}
        [1 - \rho(m)] \cdot \tau_{i, i'}(m) + \rho(m)/l(\hat{w}(m)) & \text{if } (i, i') \in \hat{w}(m),\\
        [1 - \rho(m)] \cdot \tau_{i, i'}(m)               & \text{otherwise,}
    \end{cases}    
\end{align*}
where $0 < \rho(m) < 1$ is the so-called \emph{evaporation rate}, $\hat{w}(m)$ is the best so far found path, i.e., the path with the minimum cost until the end of iteration $m$, and $l(w)$ is the length of path $w$, i.e., the number of arcs of path $w$.
To ensure that the total sum of the pheromone values remains equal to one, we normalize by setting the increment of the pheromone values at the end of cycle $m$ to $\rho(m)/l(\hat{w}(m))$.

\begin{observation}
\label{obs:sumOneGbas}
    The total pheromone value of the outgoing arcs of any node in $\mathcal{C}$ is always at most 1.
\end{observation}

In this update rule, we always reward the arcs of the best path found so far. This path is replaced each time some ant finds a path with lower costs. More precisely, if we have more than one optimal walk in $\mathcal{C}$, only one of these walks will survive and the pheromone values of the others will vanish.

\subsection{Convergence}
\label{sec:convergence}

In this section, we show that the solution generated in each iteration of GBAS/tdev converges with probability 1 to the optimal solution of the problem.
To this end, we first define the following events:
\begin{itemize}
  \item $E_m^j$ denotes the event that ant $j$ traverses an optimal walk in cycle $m$.
  \item $B_m$ denotes the event that no ant traverses an optimal walk in cycle $m$.
  \item $W$ denotes the event that at least one optimal walk is traversed by at least one ant in some cycle.
\end{itemize}

Gutjahr \cite{gutjahr2002aco} showed that the search procedure of GBAS/tdev is a \emph{Markov process}, i.e., a stochastic process where the probability distribution of the state at cycle $m$ only depends on the state at cycle $m-1$.
We provide a proof of this fact here for completeness.
Formally, let $(\underline{\tau}(m), \underline{\omega}(m), f^*(m))$ define the states of the Markov process, where $\underline{\tau}(m)$ is the vector of the pheromone values $\tau_{i, i'}(m)$ in cycle $m$ for all the arcs $(i, i') \in \mathcal{A}$,  $\underline{\omega}(m)$ is the vector of the walks $\omega^j(m)$ in cycle $m$ for all the ants $j \in S$, and $f^*(m)$ is the length of the best path found so far in cycles $1, \dotsc , m-1$ (we set $f^*(1) = \infty$).

\begin{proposition}
    The state variables $(\underline{\tau}(m), \underline{\omega}(m), f^*(m))$, for $m = 1, 2, \dotsc$, form a Markov process.
\end{proposition}
\begin{proof}
    $\underline{\tau}(m)$ is determined deterministically from $\underline{\tau}(m-1), \underline{\omega}(m-1)$ and $f^*(m-1)$ according to the pheromone update rule (the values $\rho(m)$ are deterministic).
    $\underline{\omega}(m)$ only depends on $\underline{\tau}(m)$, and therefore is determined by $\underline{\tau}(m-1), \underline{\omega}(m-1)$ and $f^*(m-1)$.
    $f^*(m)$ is determined from $\underline{\omega}(m-1)$ and $f^*(m-1)$. 
\end{proof}

To do the analysis, we first provide a lower bound for the probability that at least one ant traverses an optimal walk $\omega^* = (0, 1, \dotsc, L)$ in cycle $m$ (Lemma~\ref{lem:lowerbound1}). Next, we see that the provided lower bound holds independently of the events in the previous cycles (Corollary~\ref{cor:lowerbound2}). Finally, we prove that the probability that an optimal walk $\omega^*$ is traversed by at least one ant converges to 1.

\begin{lemma}
\label{lem:lowerbound1}
    For any cycle $m \geq 2$, we have $\probP[\neg B_m] \geq 1 - \left(1 - \left(\dfrac{\prod_{k=1}^{m-1}[1-\rho(k)]}{n^2}\right)^L\right)^{|S|}$.
\end{lemma}
\begin{proof}
    Let $\omega^* = (0, 1, \dotsc, L)$ be an optimal walk. We will provide a lower bound for the probability that at least one ant $j$ traverses $\omega^*$ in cycle $m \geq 2$. We have
    \begin{align*}
        \probP[E_m^j]  
        =  \prod_{i=0}^{L-1} p_{i, i+1}(m)
        = \prod_{i=0}^{L-1} \frac{\tau_{i, i+1}(m)}{\sum_{i': (i,i') \in \mathcal{A}} \tau_{i, i'}(m)}
        &\phantom{}\geq \prod_{i=0}^{L-1} \tau_{i, i+1}(m) \\
        &\geq \prod_{i=0}^{L-1} [1-\rho(m-1)] \cdot \tau_{i, i+1}(m-1) \\
        &\geq \prod_{i=0}^{L-1} \prod_{k=1}^{m-1} [1-\rho(k)] \cdot \tau_{i, i+1}(1) \\
        &= \prod_{i=0}^{L-1} \prod_{k=1}^{m-1} [1-\rho(k)] \cdot \frac{1}{|\A|} \\
        &\geq \left(\dfrac{\prod_{k=1}^{m-1}[1-\rho(k)]}{n^2}\right)^L.
    \end{align*}
    The first inequality holds according to \Cref{obs:sumOneGbas}. The second inequality is according to the pheromone update rule in GBAS/tdev, and the third inequality is based on the repeated application of this fact. The last inequality holds as $|\A| \leq n^2$.
    
    Since the walks of the ants are independent from each other, we get
    \begin{align*}
        \probP[B_m] \leq \left( 1 - \left(\dfrac{\prod_{k=1}^{m-1}[1-\rho(k)]}{n^2}\right)^L \right)^{|S|},
    \end{align*}
    and therefore,
    \[
        \probP[\neg B_m] \geq 1 - \left( 1 - \left(\dfrac{\prod_{k=1}^{m-1}[1-\rho(k)]}{n^2}\right)^L \right)^{|S|}. \qedhere
    \]
\end{proof}

As the provided lower bound in Lemma~\ref{lem:lowerbound1} holds independent of what happens in previous cycles, we have the following corollary.

\begin{corollary}
\label{cor:lowerbound2}
    For any cycle $m \geq 2$, we have $\probP[\neg B_m | B_1, \dotsc, B_{m-1}] \geq 1 - \left(1 - \left(\dfrac{\prod_{k=1}^{m-1}[1-\rho(k)]}{n^2}\right)^L\right)^{|S|}$.
\end{corollary}

As a result of Corollary~\ref{cor:lowerbound2}, we will compute an upper bound for the probability that none of the ants traverse the optimal walk $\omega^*$ in cycles $2, \dotsc, m$.

\begin{lemma}
\label{lem:noOptWalk}
    We have $\probP[B_2 \wedge \dotsc \wedge B_m] \leq \exp\left( \dfrac{-|S|}{n^{2L}} \sum\limits_{i=1}^{m-1} \left( \prod\limits_{k=1}^{i}[1-\rho(k)] \right)^L \right).$
\end{lemma}
\begin{proof}
According to Corollary~\ref{cor:lowerbound2}, we get
    \begin{align*}
        \probP[B_2 \wedge \dotsc \wedge B_m]
        &\phantom{}=
        \probP[B_2] \cdot \probP[B_3 | B_2] \cdot \ldots \cdot \probP[B_m | B_2 \wedge \dotsc \wedge B_{m-1}] \\
        &\leq \left(1 - \left(\dfrac{[1-\rho(1)]}{n^2}\right)^L\right)^{|S|}
        \cdot
        \left(1 - \left(\dfrac{\prod_{k=1}^{2}[1-\rho(k)]}{n^2}\right)^L\right)^{|S|} \\
        & \quad \cdot \ldots \cdot
        \left(1 - \left(\dfrac{\prod_{k=1}^{m-1}[1-\rho(k)]}{n^2}\right)^L\right)^{|S|} \\
        &= \left( \prod_{i=1}^{m-1} \left(1 - \left(\dfrac{\prod_{k=1}^{i}[1-\rho(k)]}{n^2}\right)^L\right) \right)^{|S|}.
    \end{align*}
    Moreover, because $\ln x \leq x-1$ for $0 < x < 1$, we have
    \begin{align*}
        \ln \left( \prod_{i=1}^{m-1} \left(1 - \left(\dfrac{\prod_{k=1}^{i}[1-\rho(k)]}{n^2}\right)^L\right) \right)^{|S|}
        &\phantom{}=
        |S| \sum_{i=1}^{m-1} \ln \left(1 - \left(\dfrac{\prod_{k=1}^{i}[1-\rho(k)]}{n^2}\right)^L\right) \\
        &\leq
        -|S| \sum_{i=1}^{m-1} \left(\dfrac{\prod_{k=1}^{i}[1-\rho(k)]}{n^2}\right)^L \\
        &= \dfrac{-|S|}{n^{2L}} \sum_{i=1}^{m-1} \left( \prod_{k=1}^{i}[1-\rho(k)] \right)^L,
    \end{align*}
    and therefore, 
    \[
         \left( \prod_{i=1}^{m-1} \left(1 - \left(\dfrac{\prod_{k=1}^{i}[1-\rho(k)]}{n^2}\right)^L\right) \right)^{|S|}
        \leq \exp\left( \dfrac{-|S|}{n^{2L}} \sum_{i=1}^{m-1} \left( \prod_{k=1}^{i}[1-\rho(k)] \right)^L \right). \qedhere
    \]
\end{proof}

Next, we will provide a decreasing function $\rho(k)$, such that $\rho(k)$ is the evaporation rate at cycle $k$, and we will show that the probability of finding an optimal walk $\omega^*$ converges to $1$ when $m \to \infty$.

\begin{lemma}
\label{lem:limitiEvaporation}
    We have $\lim\limits_{m \to \infty} \sum_{i=1}^{m-1} \left( \prod_{k=1}^{i}[1-\rho(k)] \right)^L = \infty$, when $\rho(k) = \dfrac{\alpha}{k}$, where
    $\alpha < \dfrac{1}{2L}$.
\end{lemma}
\begin{proof}
    Let $P_i = \prod_{k=1}^{i}\left(1-\dfrac{\alpha}{k}\right)$.
    We have $\ln P_i = \sum_{k=1}^{i} \ln \left[1-\dfrac{\alpha}{k}\right]$.
    As we have $0 \leq \dfrac{\alpha}{k} \leq 1/2$, by using the lower bound $\ln(1-x) \geq -x-x^2$ for $x \in [0, 1/2]$, we get 
    \begin{align*}
         \ln P_i \geq \sum_{k=1}^{i}\left[ -\dfrac{\alpha}{k} - \left(\dfrac{\alpha}{k}\right)^2\right] = -c \sum_{k=1}^{i} \dfrac{1}{k} -\alpha^2 \sum_{k=1}^{i} \dfrac{1}{k^2},
    \end{align*}
    and since $\sum_{k=1}^{i} \dfrac{1}{k} \leq 1 + \ln i$, we get $\ln P_i \leq -\alpha - \alpha \ln i -\alpha^2 -\alpha^2 \ln i.$
    As we have $\alpha \leq 1$, we get $\ln P_i \leq -2\alpha - 2\alpha \ln i.$
    Therefore, 
    \begin{align*}
         P_i \leq e^{-2\alpha \ln i} \cdot e^{-2\alpha} = i^{-2\alpha} \cdot e^{-2\alpha}.
    \end{align*}
    Since $\alpha L < 1/2$, $\sum_{i=1}^{\infty} \dfrac{1}{i^{2\alpha L}}$ diverges to $\infty$.
\end{proof}

\gbasmainresult*

\begin{proof}
    According to Lemma~\ref{lem:noOptWalk}, the desired probability is bounded from above by
    \begin{align*}
         \exp\left( \dfrac{-|S|}{n^{2L}} \sum\limits_{i=1}^{m-1} \left( \prod\limits_{k=1}^{i}
            \left[1-\frac{\alpha}{k}\right] \right)^L \right).
    \end{align*}
    Let $P_i = \prod_{k=1}^i \left(1 - a/k\right)$. Then we have
    \[
        \ln  P_i = \sum_{k=1}^i \ln\left(1 - \frac{\alpha}{k}\right)
            \geq \sum_{k=1}^i \left(-\frac{\alpha}{k} - \frac{\alpha^2}{k^2}\right).
    \]
    Using similar considerations as in the proof of \Cref{lem:limitiEvaporation}, we
    obtain
    \[
        \ln  P_i \geq -2\alpha\sum_{k=1}^i \frac{1}{k} \geq -2\alpha\ln i - 2\alpha.
    \]
    Thus, the desired probability is bounded by
    \[
        \exp\left(-\frac{|S|}{n^{2L}}\sum_{i=1}^{m-1} i^{-2\alpha L}e^{-2\alpha L}\right).
    \]
    Since $\alpha < \frac{1}{2L}$, we have
    $\sum_{i=1}^{m-1} i^{-2\alpha L} \geq \int_1^m i^{-2\alpha L} = \frac{1}{1-2\alpha L}m^{1-2\alpha L}$. 
    Moreover, this bound also yields $e^{2\alpha L} \leq e = O(1)$. Setting this probability
    to $\varepsilon$ and solving the resulting inequality for $m^{1-2\alpha L}$ yields the claim.
\end{proof}

Similar to Gutjahr's results on GBAS/tdev \cite{gutjahr2002aco}, \Cref{thm:gbas running time} implies
convergence to an optimal walk as $m \to \infty$. However, Gutjahr required $\rho(m) \leq 1 - \frac{\ln (m)}{\ln(m+1)}
= O\left(\frac{1}{m\ln m}\right)$. We strengthen this result to require only $\rho(m) \leq \frac{\alpha}{m}$.

\begin{corollary}\label{lem:gbas convergence limit}
    As $m \to \infty$, the probability that some ant traverses an optimal walk $\omega^*$ approaches
    1, provided the evaporation rate satisfies $\rho(m) \leq \dfrac{\alpha}{m}$ with $\alpha < \dfrac{1}{2L}$.
\end{corollary}

After an ant has traversed an optimal walk $\omega^*$, only the arcs of $w^*$ are reinforced. The pheromone on the
remaining arcs decreases monotonically.

\begin{lemma}
    Fix some $\eta > 0$. Suppose an ant traverses $\omega^*$ in cycle $m^*$. Then the pheromone value on any arc $e \in \mathcal{A} \setminus \omega^*$
    is at most $\eta$ after~$m^*/\eta^{1/\alpha}$ total cycles.
\end{lemma}

\begin{proof}
    Since $\omega^*$ is traversed in cycle $m$, we have
    \begin{align*}
        \tau_e(m^* + n) = \tau_e(m^*)\prod_{i=0}^{n-1} (1-\rho(m^* + i)) \leq \prod_{i=0}^{n-1}\left(1-\frac{\alpha}{m^*+ i}\right),
    \end{align*}
    with the inequality following from the fact that the total amount of pheromone is normalized to 1. Using the standard
    inequality $1-x \leq e^{-x}$, we bound this by
    \[
        \exp\left(-\alpha\sum_{i=1}^{n-1} \frac{1}{m^*+i}\right) \leq \exp\left(-\alpha\ln\left(\frac{m^*+n}{m^*}\right)\right)
            \leq \left(\frac{m^*+n}{m^*}\right)^{-\alpha}.
    \]
    Note that $m^* + n$ is the total number of cycles; thus, inserting $m^*/\eta^{1/\alpha}$ for $m^* + n$, we find
    $\tau_e(m^* + n) \leq \eta$ as claimed.
\end{proof}
\section{Time-Dependent Parametrization in \texorpdfstring{\boldmath$n$}{n}-ANT Algorithm}
\label{sec:nant}

\subsection{Algorithm Specification}\label{sec:mmas specification}

In this section, we consider two variants of Ant Colony Optimization which are based on the $n$-ANT algorithm: $n$-ANT/tdev ($n$-ANT with time-dependent evaporation rate) and $n$-ANT/tdlb ($n$-ANT with time-dependent lower pheromone bound).
In the $n$-ANT algorithm~\cite{Attiratanasunthron2008running}, we are given a weighted directed graph $\mathcal{G} = (\mathcal{V}, \mathcal{A})$, with $|\mathcal{V}| = n$, as well as one ant $a_i$ assigned to each node $i \in \mathcal{V}$.
Each ant $a_i$ is responsible for finding a shortest path from node $i$ to a termination node $t \in \mathcal{V}$.
A \emph{cycle} is a time period during which every ant either finds a path to $t$ or, after at most $n$ iterations, reports that such a path has not been found, since any shortest path in $\mathcal{G}$ has at most $n$ nodes.
Similar to the transition probability in GBAS, ant $a_i$ goes from node $i$ to $i'$ in cycle $m$ with probability 
\begin{align*}
    p_{i,i'}(m) = \dfrac{\tau_{i, i'}(m)}{\sum_{i'': (i, i'') \in \A} \tau_{i, i''}(m)},
\end{align*}
where $\tau_{i, i'}(m)$ is the pheromone value of the arc $(i, i')$ in cycle $m$.
At the beginning of cycle 1, we initialize $\tau_{i, i'}(1) = \frac{1}{\mathrm{deg_o}(i)}$, where $\mathrm{deg_o}(i)$ is the out-degree of node $i$. With this initialization, we ensure that initially the sum of the pheromone values of all the outgoing arcs of node $i$ is equal to 1.

In $n$-ANT/tdev, we consider a decreasing function $0 < \rho(m) < 1$ for the value of the evaporation rate at cycle $m$. 
In $n$-ANT/tdlb, the evaporation rate has a constant value $0 < \rho < 1$; however, in this variant, the pheromone values of the arcs in $\mathcal{G}$ are constrained by upper and lower bounds, with $\tau_{\max}$ as the maximum possible pheromone value and $\tau_{\min}(m)$ as the minimum possible pheromone value at cycle $m$.

Let $f_i(m)$ be the length of the path found by the ant $a_i$ from $i$ to $t$ in cycle $m$, and $\hat{f_i}$ be the length of the best path found so far by $a_i$ from $i$ to $t$.
Moreover, let $S_i(m)$ and $\hat{S_i}$ be the set of the corresponding nodes to $f_i(m)$ and $\hat{f_i}$, respectively.
At the end of each cycle $m$, first each ant $a_i$ updates the values of $\hat{f_i}$ and $\hat{S_i}$; i.e.,
if $f_i(m) < \hat{f_i}$, then we reassign $\hat{f_i} = f_i(m)$ and $\hat{S_i} = S_i(m)$.
Then, each ant $a_i$ updates the pheromone values of all outgoing arcs $(i, i')$ from $i$ based on the specific variant of the $n$-ANT algorithm being applied. 
In $n$-ANT/tdev, we have
\begin{align*}
    \tau_{i, i'}(m+1) = 
    \begin{cases}
        (1-\rho(m)) \cdot \tau_{i, i'}(m) + \rho(m) & \text{if } i' \in \hat{S_i},\\
        (1-\rho(m)) \cdot \tau_{i, i'}(m)  & \text{otherwise,}
    \end{cases}    
\end{align*}
and in $n$-ANT/tdlb, we have
\begin{align*}
    \tau_{i, i'}(m+1) = 
    \begin{cases}
        \min\{(1-\rho) \cdot \tau_{i, i'}(m) + \rho, \tau_{\max} \} & \text{if } i' \in \hat{S_i},\\
        \max\{(1-\rho) \cdot \tau_{i, i'}(m), \tau_{\min}(m) \} & \text{otherwise.}
    \end{cases}    
\end{align*}
We specifically only consider $\tau_\text{min}(m) = \frac{c_n}{\ln(m+1)}$, as recommended by Gutjahr~\cite{gutjahr2002aco}. Here, $c_n \leq \frac{1}{n^2}$.

According to these update rules, each ant $a_i$ locally updates the pheromone values of all the outgoing arcs of node $i$.
As any shortest path in $\mathcal{G}$ has at most $n$ nodes, if an ant $a_i$ does not reach $t$ after $n$ iterations in cycle $m$, we set $f_i(m) = \infty$.
We note that the update rule of the $n$-ANT/tdev has a useful normalization property.

\begin{lemma}\label{lem:total pheromone}
    The total pheromone value of the outgoing arcs of any node $i$ in $n$-ANT/tdev is always equal to 1.
\end{lemma}

\begin{proof}
    Let $\tau_{i}(m) = \sum\limits_{i':(i, i') \in \mathcal{A}}\tau_{i,i'}(m)$. By the pheromone update rule,
    exactly one arc $(i, i^*)$ gets its pheromone value increased by $\rho(m)$ in cycle $m$,
    while the remaining arcs lose some pheromone. We then have
    \begin{align*}
        \tau_{i}(m+1) 
            &= \sum_{i' \neq i^*:(i, i') \in \mathcal{A}} (1-\rho(m))\tau_{i,i'}(m) + (1-\rho(m))\tau_{i,i^*} + \rho(m) \\
            &= (1-\rho(m))\sum_{i':(i, i') \in \mathcal{A}} \tau_{i,i'}(m)
                + \rho(m).
    \end{align*}
    Since the total amount of pheromone on the outgoing arcs is one in the first cycle, the lemma
    follows inductively.
\end{proof}

Furthermore, we have the following lemma on the total amount of pheromone on the outgoing arcs of any node $i$ in $n$-ANT/tdlb.

\begin{lemma}
\label{lem:sumAtLeast1}
    The total pheromone value of the outgoing arcs of any node $i$ in $n$-ANT/tdlb is at least equal to 1 when $\tau_\text{max} \geq 1$.
\end{lemma}

\begin{proof}
    Initially, the sum of the pheromone values of all the outgoing arcs of node $i$ is equal to 1 as $\tau_{i, i'}(1) = \frac{1}{\mathrm{deg_o}(i)}$.
    During the pheromone update in cycle $m$, if no arc's pheromone value exceeds the upper bound $\tau_\text{max}$ or falls below the lower bound $\tau_\text{min}(m)$, the sum of the pheromone values of all the outgoing arcs of node $i$ remains unchanged. 
    As we have $\tau_\text{max} \geq 1$, if the pheromone value on at least one arc gets decreased to  $\tau_\text{max}$, or gets increased to  $\tau_\text{min}(m)$, the sum is at least equal to 1.
\end{proof}

We now proceed to analyze the running time of $n$-ANT/tdev and $n$-ANT/tdlb. For both of these algorithms,
we derive a recurrence relation for the time that an ant $a_i$ takes to find its shortest path,
as a function of the time taken for ants ``closer'' to the target node. In the case of $n$-ANT/tdev,
the recurrence relation yields a lower bound for the running time, while for $n$-ANT/tdlb
we obtain an upper bound.

\subsection{Running Time Analysis of \texorpdfstring{{\boldmath$n$}}{n}-ANT/tdev}

\begin{figure}
    \centering
    \begin{tikzpicture}[scale=1.2]
        \begin{scope}[
                   decoration={
                       markings,
                       mark=at position 0.5 with {\arrow{>}}}
               ]   
        \vertex[label=below:$4$] (v1) at (-5, 0) {};
        \vertex[label=below:$3$] (v2) at (-4, 0) {};
        \vertex[label=below:$2$] (v3) at (-3, 0) {};
        \vertex[label=below:$1$] (v4) at (-2, 0) {};
        \vertex (v5) at (-1, 0) {};
        \vertex[label=right:$t$] (t) at (0, 0) {}; 

        \foreach \i in {1,...,4} {
            \ifthenelse{\i<4}{
                \draw[postaction=decorate][evaluate={\j=\i+1;}] (v\i) -- (v\j);
            }{};
            \draw[postaction=decorate] (v\i) to[out=50,in=130] (v5);
        }
        \draw[postaction=decorate] (v5) -- node[above] {$M-1$} ++ (t);
        \draw[postaction=decorate] (v4) to[out=-50,in=-130] (t);
        \end{scope}
    \end{tikzpicture}
    \caption{The series instance for $n = 4$. Arc weights not shown are equal to 1.
    The first four nodes on the left form the chain $C$; the remaining unlabeled
    node is the attractor node.}
    \label{fig:series}
\end{figure}
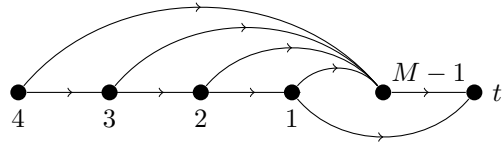

In this part, we provide a super-polynomial lower bound on the total number of cycles required for all ants $a_i$ to find a shortest path from their respective starting nodes $i$ to the target node $t$.
To this end, we consider an instance with a directed path $P$ of $n+2$ nodes, along with some extra arcs. The last node on this path
is the target node $t$. We call the node preceding
the target node the \emph{attractor node}.
The remaining $n$ nodes form the \emph{chain} $C$, are labeled as $\{1, \ldots, n\}$. Node $1$ has two outgoing arcs: one leading to the attractor node, which we call its \emph{incorrect arc}, and one directly to the target node, which we refer to as its \emph{correct arc}. For each subsequent node $i > 1$, there are also two outgoing arcs: one to the attractor node (its \emph{incorrect arc}) and one to node $i - 1$ (its \emph{correct arc}). Every arc in this instance has a weight of $1$, except the arc from the attractor node to the target node, which has a weight of $M-1$, where $M > n$. We call this structure the \emph{series} instance (see Fig.~\ref{fig:series}).

In the series instance, each node $i$ in the chain $C$ is associated with an ant $a_i$.
In order to find the shortest path, each ant on node $i$ in $C$ must always choose the correct arc of this node. If an ant chooses an incorrect arc at any point, then it finds a strictly longer path due to the final arc from the attractor to the target node.

For the analysis, we assume that $\rho(m) = \alpha_n/m^\beta$ with $\alpha_n < 1$ and $\beta \geq 0$. 
Here, $a_n$ in general depends
on $n$; however, we assume $\beta$ is a fixed constant. Moreover, we assume that if $\alpha_n$ depends
on $n$, then it is a non-increasing function of $n$. Note that this is also the case in \Cref{thm:gbas running time}.
We divide the analysis into two cases, depending on the value of $\beta$.
It turns out that the case of $\beta > 1$ has an easy analysis compared to the case of $\beta \leq 1$.

\subsubsection{The Easy Case: \texorpdfstring{$\beta > 1$}{beta > 1}}
For the analysis of this case, we first bound the probability that an ant makes the correct decision in any iteration.

\begin{lemma}\label{lem:chain small prob good choice}
    In every cycle, the probability that an ant starting from node $i$ in the chain chooses the correct arc of this node in any iteration of this cycle is bounded from above by a constant
    $p < 1$.
\end{lemma}

\begin{proof}
    Fix some ant to consider for this lemma; its starting point is irrelevant.
    Suppose the ant is at node $i \in C$ in this iteration. Let the amount
    of pheromone on the arc from $i$ to $i-1$ in cycle $m$ be $\tau_\text{correct}$, and define
    $\tau_\text{incorrect}$ as the pheromone on the other arc from $i$. By \Cref{lem:total pheromone}, 
    we have $\tau_\text{correct} = 1 - \tau_\text{incorrect}$, and
    the probability that the ant chooses this arc is equal to $\tau_\text{correct}$.
    
    By the pheromone update rule, we have
    $\tau_\text{incorrect} \geq \frac{1}{2}\prod_{t=1}^{m-1} (1-\rho(t)) 
    \geq \frac{1}{2}\prod_{t=1}^\infty (1 - \rho(t))$. 
    With $\rho(t) = \alpha_n t^{-\beta}$, we have
        $\prod_{t=1}^\infty (1 - \rho(t)) = \prod_{t=1}^\infty (1 - \alpha_n t^{-\beta})$.
    Since $\beta > 1$ and $\alpha_n < 1$ this product evaluates to $\Omega(1)$. Hence, the desired
    probability is $1 - \Omega(1)$ and is thus bounded from above by a constant
    strictly smaller than 1, for every cycle.
\end{proof}

Using this lemma, we easily get our result.

\begin{lemma}\label{lem: running time large b}
    The number of cycles $n$-ANT/tdev using $\rho(m) = \alpha_n/m^\beta$ with $\alpha_n < 1$ and $\beta > 1$ 
    takes on the series instance 
    for all ants to see their shortest paths is $2^{\Omega(n)}$ in
    expectation.
\end{lemma}

\begin{proof}
    Consider the ant that starts at node $n$. The running time of $n$-ANT/tdev is bounded
    from below by the first cycle number in which this ant sees its shortest path.
    To see its shortest path, this ant needs to pick the correct arc $n$ times in
    a row. The probability that the ant does this in any particular cycle is bounded by
    $p^{n}$ for some constant $p < 1$, 
    which follows from \Cref{lem:chain small prob good choice} and the
    fact that the choice the ant makes in any iteration only depends on its current
    position. 
    Consequently, the number of cycles until this ant first sees its shortest path
    stochastically dominates a geometric random variable with parameter $p^n$,
    and so the expected number of cycles is at least 
    $\left(\frac 1p\right)^n = 2^{\Omega(n)}$.
\end{proof}

\subsubsection{The Hard Case: \texorpdfstring{$\beta \leq 1$}{beta <= 1}}

For the case $\beta \leq 1$, the previous analysis fails,
since now $\prod_{i=1}^\infty (1-\rho(i)) = 0$. We must now carefully
consider the evolution of the pheromone values during the execution of the algorithm,
as we cannot assume that there is always a constant probability of any ant making the wrong choice
during any iteration in a cycle.

First, we need a few definitions. 
We say that a node $i \in C$ is \emph{$\varepsilon$-processed} if the incorrect arc of $i$ has a pheromone value of at most $\varepsilon \in (0, 1)$, which is a fixed constant that we leave
ambiguous -- its precise value is not important. For most of this subsection, we simply say \emph{processed}
for brevity, since the value of $\varepsilon$ does not change throughout the analysis.
For an ant $a_i$, we define
$L_i(m)$ to be the number of ants with label less than $i$ that have \emph{not}
processed their node
in cycle $m$. Note that $L_i(m)$ is a random variable. 

As each non-processed node has a pheromone value of at most $(1- \varepsilon)$
on its good arc, the probability
that an ant $a_i$ finds its shortest path in cycle $m$ is at most $(1-\varepsilon)^{L_i(m)}$, since
it needs to choose the correct arc on each node with a label smaller than $i$. This immediately yields
the following.

\begin{lemma}\label{lem: find shortest path worst case}
    The probability that $a_i$ finds its shortest path in cycle $m$, given that $L_i(m) \geq c\sqrt{n}$ for some
    constant $c > 0$, is bounded
    from above by $2^{-\Omega(\sqrt{n})}$.
\end{lemma}


This shows that, if $\Omega(\sqrt n)$ nodes ahead of some ant $a_i$ are not processed, then $a_i$ is unlikely to find
its shortest path in $2^{o(\sqrt{n})}$ cycles.

Next, we need a lower bound on the number of cycles needed for an ant to process its node.

\begin{lemma}\label{lem: time to process worst case}
    Suppose $a_i$ sees its optimal solution for the first time in cycle $m$, and it has never reinforced its
    correct arc before cycle $m > 1$. Then the number of cycles that $a_i$ needs to process node $i$ is at least
    \begin{enumerate}
        \item $\left(\left(\frac{1}{2\varepsilon}\right)^{\frac{1-\alpha_n}{\alpha_n}}-1\right) \cdot m + 1 - \left(\frac{1}{2\varepsilon}\right)^{\frac{1-\alpha_n}{\alpha_n}}$ cycles, if $\beta = 1$;
        \item $\left((m-1)^{1-\beta} + (1-\beta)\cdot\frac{1-\alpha_n}{\alpha_n}\ln\frac{1}{2\varepsilon}\right)^{\frac{1}{1-\beta}} - (m-1) \geq
        \left((1-\beta) \cdot \frac{1-\alpha_n}{\alpha_n}\ln\frac{1}{2\varepsilon}\right)^{\frac{1}{1-\beta}}$ cycles, if $\beta < 1$.
    \end{enumerate}
\end{lemma}

\begin{proof}
    Let $e = (i, i')$ be the incorrect arc of node $i$.
    Since $a_i$ never reinforced its correct arc before $m$, the pheromone value on $e$ never decreased before $m$;
    hence, $\tau_e(m) \geq \frac 12$.
    So in $\delta$ cycles after $m$ we have
    \[
    \tau_e(m+\delta) \geq \frac 12 \prod_{t=m}^{m+\delta-1}\left(1-\dfrac{1}{t^\beta}\right)
    = \frac 12 \prod_{k=0}^{\delta-1}\left(1-\dfrac{\alpha_n}{(m+k)^\beta}\right).
    \]
    We want to ensure that node $i$ is not processed before $\delta$ cycles have passed. It then suffices
    to bound the product above from below by $\varepsilon$:
    \[
     \frac 12 \prod_{k=0}^{\delta-1}\left(1-\dfrac{\alpha_n}{(m+k)^\beta}\right) \geq \varepsilon,
    \]
    which is equivalent to 
    \[
    \sum_{k=0}^{\delta-1} \ln{\left( 1-\dfrac{\alpha_n}{(m+k)^\beta} \right)} \geq \ln{\varepsilon} + \ln 2 = \ln(2\varepsilon).
    \]
    As $\ln{(1+x)} \geq \frac{x}{1+x}$ for $x > -1$, it suffices to determine $\delta$ such that
    \[
    \sum_{k=0}^{\delta-1}\frac{\alpha_n}{(m + k)^\beta} \cdot \frac{1}{1 - \frac{\alpha_n}{(m+k)^\beta}} \leq \ln\frac{1}{2\varepsilon}.
    \]
    Since $m \geq 1$ and $\alpha_n < 1$, we have $\frac{1}{1-\frac{\alpha_n}{(m+k)^\beta}} \leq \frac{1}{1-\alpha_n}$. Thus, it in turn
    suffices to compute the largest $\delta$ such that
    \[
    \sum_{k=0}^{\delta-1}\frac{\alpha_n}{(m + k)^\beta} \leq (1-\alpha_n)\ln\frac{1}{2\varepsilon}.
    \]
    Since $a\sum_{k=0}^{\delta-1}\dfrac{1}{(m+k)^\beta}  = \alpha_n\sum_{k=m}^{m+\delta-1}\frac{1}{k^\beta} \leq \alpha_n\int_{z=m-1}^{m+\delta-1} \dfrac{1}{z^\beta} \dd z$ for $m > 1$, we get
    \begin{align}
    \label{eq:expLowerboundInt}
        \alpha_n\int_{m}^{m+\delta-1} \dfrac{1}{z^\beta} \dd z \leq (1-\alpha_n)\ln\frac{1}{2\varepsilon}.
    \end{align}
    
    We have two cases based on the value of $\beta$. 
    For $\beta=1$, we have
    \[
    \int_{m-1}^{m+\delta-1} \dfrac{1}{z^\beta} \dd z = \ln{(m+\delta-1)} - \ln{(m)}.
    \]
    Therefore, according to \Cref{eq:expLowerboundInt}, we get
    \[
    \ln\left({\dfrac{m+\delta-1}{m-1}}\right) \leq \frac{1-\alpha_n}{\alpha_n}\ln\frac{1}{2\varepsilon},
    \]
    which is equivalent to
    \[
    \delta \leq \left(\left(\frac{1}{2\varepsilon}\right)^{\frac{1-\alpha_n}{\alpha_n}}-1\right) \cdot m
        + 1 - \left(\frac{1}{2\varepsilon}\right)^{\frac{1-\alpha_n}{\alpha_n}}. \qedhere
    \]
    This proves the first part of the lemma.
    
    For $\beta<1$, we have,
    \[
    \int_{m-1}^{m+\delta-1} \dfrac{1}{z^\beta} \dd z = \dfrac{(m+\delta-1)^{1-\beta} - (m-1)^{1-\beta}}{1-\beta}.
    \]
    Therefore, according to \Cref{eq:expLowerboundInt}, we get
    \[
    \dfrac{(m+\delta-1)^{1-\beta} - (m-1)^{1-\beta}}{1-\beta} \leq \frac{1-\alpha_n}{\alpha_n}\ln\frac{1}{2\varepsilon}.
    \]
    Solving this for $\delta$ yields the second part of the lemma.
\end{proof}

For the next step, we divide the ants starting on the chain $C$ into $k = \ceil{\sqrt{n}}$ 
segments $S_i$, each of size of at least $\floor{\sqrt n}-1$. 
For a fixed constant $\eta \in (0, 1)$, we say that a segment $S_i$ is $\eta$-processed if at least $\eta|S_i|$
ants in $S_i$ have processed their nodes. As with $\varepsilon$, we leave the value of $\eta$ ambiguous,
as its precise value is not important.
Let $T_i$ be the first cycle in which segment $S_i$ is $\eta$-processed.
In the following, we write \emph{with high probability} or
\whp{} as a shorthand for
``with probability $1 - 2^{-\Omega(\sqrt n)}$''.
We need a simple Chernoff bound.

\begin{lemma}[Chernoff Bound \cite{mitzenmacherUpfal2005}]\label{lem: chernoff}
    Let $X_1,\ldots,X_k$ be a sequence of independent random variables taking
    values in $\{0, 1\}$. Let $X = \sum_{i=1}^k X_i$ and $\mu = \exP[X]$. Then
    for any $\delta > 0$,
    \[
        \probP\left(X > (1 + \delta)\mu\right) \leq e^{-\delta^2\mu/(2 + \delta)}
    \]
    and for any $0 < \delta < 1$
    \[
        \probP\left(X < (1 - \delta)\mu\right) \leq e^{-\delta^2\mu/2}.
    \]
\end{lemma}

We wish to show that the variables $T_i$ satisfy a certain recurrence inequality with
high probability. 
For this, we would like to use \Cref{lem: time to process worst case}, which assumes
that an ant $a$ has never reinforced its correct arc. \Cref{lem: never reinforce correct arc}
ensures that this holds for a constant fraction of ants in each segment $S_i$ \whp{}.


%
%
%

\begin{lemma}\label{lem: never reinforce correct arc}
    With high probability, every segment $S_i$ contains $\Omega(\sqrt n)$ ants
    that never reinforce their correct arcs until they first see their shortest paths.
\end{lemma}

\begin{proof}
    
    First, note that an ant $a_j \in S_i$ has two possibilities: it either
    chooses the correct arc, or it chooses the incorrect arc, leading to a path of weight $M$.
    If the ant chooses the correct arc in the first iteration of any cycle,
    but does not find its shortest path, then it must choose an incorrect arc in a later iteration
    of the same cycle.
    This leads to a path of length at least $M+1$.
    
    Thus if the ant chooses the incorrect arc from $j$, then it finds a strictly shorter path than if it
    chooses first the correct arc and later an incorrect arc.
    Consequently, if the ant chooses the incorrect arc in the first iteration of the first cycle, 
    it never reinforces the correct arc until it first sees its shortest path.
    This follows because the pheromone update rule only reinforces the arc that leads to the best path found so far.

    In the first cycle, each ant in $S_i$ flips an unbiased coin to determine whether
    to take the correct arc or the incorrect arc. \Cref{lem: chernoff} implies that the number
    of ants that choose the incorrect arc in the first cycle is at least $\frac 12 (1 - \delta)|S_i|$
    \whp{}.
    Thus, $S_i$ contains
    $\Omega(\sqrt n)$ ants that choose the incorrect arc in the first cycle.
    By taking a union bound over
    each of the $\ceil{\sqrt n}$ segments, we see that this holds for all segments
    simultaneously \whp{}.
\end{proof}

We are now in a position to derive the required recurrence inequality.

\begin{lemma}\label{lem: recurrence worst case}
    Let $a \in S_i$ be an ant that has not reinforced its correct arc before first seeing
    its shortest path.
    Let $F(m)$ denote a \whp{} lower bound for the number of cycles needed for $a$
    to first see its shortest path, provided that $S_{i-1}$ is first $\eta$-processed in cycle $m$.
    Let $G(m)$ denote a \whp{} lower bound for the number of cycles needed for $a$ 
    to process its node, provided that it first sees its shortest path in cycle $m$.
    Then either $T_i = 2^{\Omega(\sqrt n)}$ for some $i \in [k]$ for some constant $c > 0$, 
    or $T_i$ satisfies \whp{} the recurrence inequality $T_i \geq T_{i-1} + F(T_{i-1}) + G(T_{i-1} + F(T_{i-1}))$.
\end{lemma}

\begin{proof}
    Suppose that $T_i = 2^{o(\sqrt n)}$ for each $i \in [k]$. Then by \Cref{lem: find shortest path worst case},
    the probability that an ant in $S_i$ finds its shortest path before $T_{i-1}$ is $2^{-\Omega(\sqrt n)}$,
    since for any ant $a \in S_i$ starting at node $j$ we have $L_j(m) = \Omega(\sqrt n)$ for any $m < T_{i-1}$.
    If we furthermore assume that $a$ has not reinforced its correct arc
    path before seeing its shortest path, then $a$ needs \whp{}
    $F(T_{i-1})$ cycles after $T_{i-1}$ to find its shortest path. Subsequently,
    $a$ needs $G(T_{i-1} + F(T_{i-1}))$ cycles to process its node.

    If we could guarantee that no ant in $a \in S_i$ reinforces its correct arc before seeing its shortest path,
    we could immediately say that 
    $T_i \geq T_{i-1} + F(T_{i-1}) + G(T_{i-1} + F(T_{i-1}))$ \whp{}.
    In reality, some ants may reinforce their correct arcs before $T_{i-1}$, so we must deal with this event. However, \Cref{lem: never reinforce correct arc} guarantees that 
    \whp{} every segment contains $\Omega(\sqrt n)$ ants that
    never reinforce their correct arc before first finding their shortest paths,
    which suffices for our purposes.
    Consequently, every segment contains $\Omega(\sqrt n)$ ants for which the assumptions on the ant $a$
    from the lemma statement hold \whp{},
    and so the sequence $T_i$ indeed satisfies the claimed recurrence \whp{}.
\end{proof}

From \Cref{lem: recurrence worst case} we obtain a running time lower bound for $b = 1$
straightforwardly by solving the recurrence inequality.

\begin{lemma}\label{lem: running time medium b}
    The number of cycles $n$-ANT/tdev using $\rho(m) = \alpha_n/m$ with $\alpha_n < 1$
    takes on the series instance
    for all ants to find their shortest paths is $2^{\Omega(\sqrt n)}$ in
    expectation.
\end{lemma}

\begin{proof}
    Let $k = \ceil{\sqrt n}$. For each ant in $S_{k}$ to find their shortest paths, we require $S_{k - 1}$ to be processed. Thus, $\exP[T_{k-1}]$ is a lower bound for the 
    expected number of cycles after which all ants have seen their optimal solution.
    By \Cref{lem: recurrence worst case} the variables $T_{i}$
    satisfy for each $2 \leq i \leq k$ \whp{} the recurrence inequality
    \begin{align*}
        T_{i} &\geq
                T_{i-1} + F(T_{i-1}) + G(T_{i-1} + F(T_{i-1})) \\
                &\geq 
                T_{i-1} + F(T_{i-1}) + \left\lceil T_{i-1} \cdot \left(
                \left(\frac{1}{2\varepsilon}\right)^{\frac{1-\alpha_n}{\alpha_n}} - 1\right) + 1 -\left(\frac{1}{2\varepsilon}\right)^{\frac{1-\alpha_n}{\alpha_n}}\right] \\
            &\geq c_1\cdot T_{i-1} + c_2.
    \end{align*}
    Here we used $G(m)$ according to \Cref{lem: time to process worst case}. 
    If we choose $\varepsilon$ sufficiently small, then this recurrence inequality is solved by
    $T_{k-1} = 2^{\Omega(k)} = 2^{\Omega(\sqrt n)}$ \whp{}.
    Consequently, the same bound holds for $T_{k-1}$ in expectation, which yields the claim.
\end{proof}

In the proof of \Cref{lem: running time medium b}, we could ignore the term in the recurrence
inequality that involves $F(\cdot)$, since for $\beta = 1$ the function $G(m)$ is proportional to $m$.
However, for $\beta < 1$ we instead have $G(m) \propto m^\beta$, which is not sufficient on its own
guarantee a super-polynomial running time. This complicates the analysis significantly. Hence,
we need to lay some more groundwork to proceed.

The following lemma says that for $\beta < 1$, the ants
in each segment need many trials before they first see their shortest paths.

\begin{lemma}\label{lem: time to find worst case}
    Suppose $\beta < 1$, and assume $S_{i-1}$ is first processed in cycle $m$.
    Then \whp{}, there are $\Omega(\sqrt n)$ ants in
    $S_i$ that need at least
    \[
        \exp\left(\frac{a}{1-\beta}\left(m^{1-\beta} - 1\right)\right)
    \]
    cycles after $m$ to see their shortest paths for the first time.
\end{lemma}

\begin{proof}
    From \Cref{lem: never reinforce correct arc}
    we know that \whp{} there are $\Omega(\sqrt n)$ ants in $S_i$ that have not seen their shortest paths before
    cycle $m$, and consequently never reinforced their correct arcs before cycle $m$.
    Let $S_i' \subseteq S_i$ denote the ants in $S_i$ satisfying these premises.
    
    Consider an ant $a \in S_i'$. The amount of pheromone on its correct arc at the start of cycle $m$ is
    \begin{align}\label{eq: success prob}
        \frac 12 \prod_{t=1}^{m-1} \left(1-\frac{\alpha_n}{t^\beta}\right)
            &\leq \frac 12 \exp\left(-\alpha_n\sum_{t=1}^{m-1} t^{-\beta}\right)
            &\leq \frac 12 \exp\left(-\alpha_n\int_{1}^{m} t^{-\beta}\dd t\right) \nonumber \\
            &= \frac 12 \exp\left(-\frac{\alpha_n}{1-\beta}\left(m^{1-\beta} - 1\right)\right).
    \end{align}
    Moreover, since this function is
    decreasing in $m$, the bound decreases each time $a$ fails to find
    its shortest path.
    
    To find its shortest path, $a$ must first choose its correct arc. Hence, the number
    of ants that find their shortest paths in cycle $m$ is bounded from above by the number of ants
    that choose their correct arcs in this cycle.  We only consider the latter
    quantity, as it is much simpler to analyze and suffices for our purposes.

    Let $p$ denote the bound in \Cref{eq: success prob}. Then the number of
    trials that $a$ needs until it succeeds stochastically dominates a geometric
    random variable with parameter $p$. Hence, the probability that $a$ needs $\ell$
    or fewer trials is at most $1 - (1-p)^\ell$. Using Bernouilli's inequality,
    this is bounded from above by $\ell p$. Therefore, the expected number of ants
    in $S_i'$ that choose their correct arcs in $\ell$ or fewer trials is
    at most
    \[
        \ell p|S_i'| = \frac{\ell}{2}\exp\left(-\frac{\alpha_n}{1-\beta}\left(m^{1-\beta} - 1\right)\right)|S_i'|
        \leq \frac 12 |S_i'|,
    \]
    where in the last inequality we set 
    $\ell = \exp\left(\dfrac{\alpha_n}{1-\beta}\left(m^{1-\beta}-1\right)\right)$.

    We now find that the expected 
    number of ants that need $\ell$ or more trials to choose their correct arcs
    is at least $\frac 12 |S_i'|$. To turn this into a statement that holds \whp{},
    we define for each $a \in S_i'$ an indicator variable $X_a$, taking a value
    of $1$ if $a$ fails to choose its correct arc in $\ell$ or fewer trials
    and $0$ otherwise. Since the ants act independently in the first iteration of 
    any cycle, the variables $X_a$ are mutually independent, and we can use
    \Cref{lem: chernoff}. Recalling that $|S_i'| = \Omega(\sqrt n)$ \whp{}
    concludes the proof.
\end{proof}

As in the proof \Cref{lem: running time medium b}, we need to solve a recurrence relation
obtained from \Cref{lem: recurrence worst case}. The following lemma helps us in a crucial step
in solving this recurrence.

\begin{restatable}{lemma}{hardrecursion}
\label{lem: hard recursion for lower bound}
    Let $0 < b < 1$ and let $c > 0$. Suppose the sequence $(M_k)_{k\geq 1}$ satisfies the recurrence 
    relation
    $M_k = M_{k-1} + c \cdot (M_{k-1})^{b}$ for $k \geq 2$
    with initial condition $M_1 = (1-b)^{\frac{1}{1-b}}c^{\frac{1}{1-b}}$.
    Then
    \[
        M_k \geq \left[c\cdot (1-b)\right]^{\frac{1}{1-b}} \cdot \left[1 + \gamma_b\cdot (k-1)\right]^{\frac{1}{1-b}},
    \]
    where $\gamma_b$ is the unique solution in $(0, 1)$ of the equation $x + x^{\frac{b}{1-b}} = 1$,
    and thus depends only on $b$.
\end{restatable}

All that is needed now to proceed to the last result of this section is one last technical lemma.

\begin{restatable}{lemma}{lowerboundbsmall}
\label{lem:lowerBoundForBLessThan1}
    For $x > 0, a > 0$ and $ 0 < b < 1$ it holds that
    \[
    \left(x^{b} + a\right)^{\frac{1}{b}} - x \geq \frac{a}{b} \cdot x^{1-b}.
    \]
\end{restatable}

This last technical lemma now finally puts us in a position to prove a running time bound for
$n$-ANT/tdev on the series instance when $\beta < 1$.

\begin{lemma}\label{lem: running time small b}
    The number of cycles $n$-ANT/tdev using $\rho(m) = \alpha_n/m^\beta$ with $\alpha_n < 1$
    and $\beta < 1$
    takes on the series instance
    for all ants to find their shortest paths is $2^{\Omega\left(\sqrt{n}\right)}$ in
    expectation.
\end{lemma}

\begin{proof}
    Let $k = \ceil{\sqrt n}$. Similar to the proof of Lemma~\ref{lem: running time medium b}, we compute a lower bound for $\exP[T_{k-1}]$. 
    According to \Cref{lem: time to process worst case,lem: recurrence worst case,lem: time to find worst case}
    we have for each $2 \leq i \leq k$
    \begin{multline*}
        T_{i} \geq 
        T_{i-1} + \exp\left(\frac{\alpha_n}{1-\beta}\left((T_{i-1})^{1-\beta} - 1\right)\right) \\
            + \left( (T_{i-1} -1)^{1-\beta} + (1-\beta)\cdot\frac{1-\alpha_n}{\alpha_n}\ln{\frac{1}{2\varepsilon}} \right)^{\frac{1}{1-\beta}} - (T_{i-1} - 1)
    \end{multline*}
    which according to \Cref{lem:lowerBoundForBLessThan1} implies
    \begin{align}
        T_i \geq T_{i-1} + \exp\left(\frac{\alpha_n}{1-\beta}\left((T_{i-1})^{1-\beta} - 1\right)\right)
            + \left(\frac{1-\alpha_n}{\alpha_n} \ln{\frac{1}{2\varepsilon}}\right) \cdot (T_{i-1}-1)^\beta. \label{eq: recurrence small b}
    \end{align}
    
    According to \Cref{lem: time to find worst case}, we also 
    have $T_1 \geq 1 + (1-\beta)^{\frac{1}{1-\beta}}\left(\frac{1-\alpha_n}{\alpha_n}\ln \frac{1}{2\varepsilon}\right)^{\frac{1}{1-\beta}}$; this holds since
    \whp{} there are $\Omega(\sqrt n)$ ants in $S_1$ that need at least two cycles to find their shortest paths (this follows easily from
    \Cref{lem: chernoff}), and hence the
    second part of
    \Cref{lem: time to process worst case} applies to these ants.

    If we ignore the second term above, then we obtain the simpler recurrence inequality
    \begin{align*}
        T_{i} \geq T_{i-1} 
            + \left(\frac{1-\alpha_n}{\alpha_n} \ln\frac{1}{2\varepsilon}\right) \cdot (T_{i-1} - 1)^\beta, 
                \quad T_1 \geq 1 + (1-\beta)^{\frac{1}{1-\beta}}\left(\frac{1-\alpha_n}{\alpha_n} \ln \frac{1}{2\varepsilon}\right)^{\frac{1}{1-\beta}}.
    \end{align*}
    This is almost in the form required by \Cref{lem: hard recursion for lower bound}. We apply that lemma
    instead to the sequence $(T_i-1)_{i=1}^{k}$, which satisfies a recurrence inequality of the
    correct form. We then have
    for each $1 \leq i \leq k$,
    \begin{align}\label{eq: solution recurrence  small b}
        T_i \geq 1 + (1-\beta)^{\frac{1}{1-\beta}}\left(\frac{1-\alpha_n}{\alpha_n}\ln \frac{1}{2\varepsilon}\right)^{\frac{1}{1-\beta}}
            \cdot \left[1 + \gamma_\beta\cdot (i-1)\right]^{\frac{1}{1-\beta}},
    \end{align}
    where $\gamma_\beta$ depends only on $\beta$.

    In \Cref{eq: recurrence small b}, let $i = k-1$. Since each term in this inequality is non-negative,
    we can take for a lower bound only the second term. Plugging in \Cref{eq: solution recurrence  small b}
    then yields
    \begin{align*}
        T_{k-1} &\geq \exp\left(
            \frac{\alpha_n}{1-\beta} \cdot \left((1-\beta)\cdot\frac{1-\alpha_n}{\alpha_n} \ln\left(\frac{1}{2\varepsilon}\right) \cdot \gamma_\beta \cdot (k-2) - 1\right)
        \right) \\
        &= e^{-\frac{\alpha_n}{1-\beta}} \cdot e^{\gamma_\beta \cdot (1-\alpha_n)\cdot \ln\left(\frac{1}{2\varepsilon}\right) \cdot (k-2)}.
    \end{align*}
    
    Note that $\beta$ and $\varepsilon$ are both constants. Moreover, we require $\alpha_n < 1$ with $\alpha_n$ non-increasing
    in $n$. Thus, we have $1 - \alpha_n = \Omega(1)$.
    Finally, recalling that $k = \Omega(\sqrt n)$, we obtain $T_{k-1} = 2^{\Omega(\sqrt n)}$ \whp{}
    and in expectation.
\end{proof}

\Cref{lem: running time small b,lem: running time medium b,lem: running time large b} yield the main result of
this section.

\nantrunningtimetdev*

\subsection{Running Time Analysis of \texorpdfstring{\boldmath$n$}{n}-ANT/tdlb}

In this part, we analyze the total number of cycles required for all ants $a_i$ to find a shortest path from their respective starting nodes $i$ to the target node $t$.
We say an ant $a_i$ \emph{sees} its optimal solution if $\hat{f_i} = f^*_i$, where $f^*_i$ is the length of the shortest path from node $i$ to $t$.
Let $S^*_i$ be the set of the corresponding nodes to $f^*_i$.
For an ant $a_i$, an arc $(i, i')$ is called \emph{incorrect} if $i' \notin S^*_i$.
Moreover, we say that node $i$ is \emph{$\varepsilon$-processed} if $\tau_{i, i'} \leq \varepsilon = 1/n^2$ for all the incorrect arcs $(i, i')$.
Let $1, 2, \dotsc, n=t$ be an enumeration of the vertices in $\mathcal{G}$ with $\ell(1) \geq \ell(2) \geq \dotsc \geq \ell(n)$ where $\ell(i)$ is the maximum number of arcs on any shortest path from node $i$ to $t$. With this ordering of the vertices, all the shortest paths from node $i$ to $t$ only use vertices from $\{ i+1, i+2, \dotsc, n \}$; otherwise, there would exist a shortest path from $i$ to $t$ with more than $\ell(i)$ arcs.

To analyze the total number of cycles, we first derive an upper bound on the number of cycles required for node $i$ to become $\varepsilon$-processed after ant $a_i$ sees its optimal solution for the first time (Lemma~\ref{lem:getProcessed1}). 
Then, assuming that all nodes reachable from $i$ to $t$ have already been $\varepsilon$-processed, we compute the expected number of cycles until ant $a_i$ sees its optimal solution (Lemma~\ref{lem:firstTimeSeeOpt}).
To determine this expected value, we first establish a lower bound on the probability that ant $a_i$ sees its optimal solution in any cycle $m$ under the same assumption (Lemma~\ref{lem:probOfSeeingOpt}). 
Finally, in Theorem~\ref{thm:totalCycles}, we report the total number of cycles for all ants to find
their shortest paths to the target node.

\begin{lemma}
\label{lem:getProcessed1}
    Suppose ant $a_i$ sees its optimal solution for the first time in cycle $m^*$. In at most $O\left(\frac{1}{\rho} \ln \dfrac{\tau_\text{max}}{\varepsilon}\right)$ cycles after $m^*$, node $i$ is $\varepsilon$-processed.
\end{lemma}

\begin{proof}
    Consider any incorrect arc $e = (i, i')$. Let $\tau$ be the pheromone value of $e$ at the beginning of cycle $m^* + 1$.
    Since $a_i$ has seen its optimal solution in cycle $m^*$, the pheromone value of $e$ decreases by a factor of $(1 - \rho)$ in every cycle after $m^*$.
    Let $\theta = \frac{1}{\rho} \ln \dfrac{\tau_\text{max}}{\varepsilon}$. The pheromone value of $e$ in cycle $m^* + \theta$ is
    
    \[
        \tau \cdot (1-\rho)^{\frac{1}{\rho} \ln \dfrac{\tau_\text{max}}{\varepsilon}}
        \leq \tau_\text{max} \cdot (1-\rho)^{\frac{1}{\rho} \ln \dfrac{\tau_\text{max}}{\varepsilon}}
        \leq \tau_\text{max} \cdot (e^{-\rho})^{\frac{1}{\rho} \ln \dfrac{\tau_\text{max}}{\varepsilon}}
        = \varepsilon.
        \qedhere
    \]
\end{proof}

Next, we bound the expected number of cycles to $\varepsilon$-process node $i$, given that all the nodes reachable from $i$ to $t$ have already been $\varepsilon$-processed. Note that since we have $\ell(1) \geq \ell(2) \geq \dotsc \geq \ell(n)$, all the nodes reachable from $i$ to $t$ are from the set $\{ i+1, i+2, \dotsc, n \}$.

\begin{lemma}
\label{lem:probOfSeeingOpt}
    The probability that ant $a_i$ sees its optimal solution in cycle $m$, given that all the nodes reachable from $i$ to $t$ have already been $\varepsilon$-processed, is at least $\Omega\left(\dfrac{\tau_\text{min}(m)}{n \cdot \tau_\text{max}}\right)$ when $\tau_\text{max} \geq 1$.
\end{lemma}

\begin{proof}
 An ant $a_i$ has to make at most $n$ random choices from node $i$ to $t$. In cycle $m$, it goes correctly from node $i$ to $i' \in \{ i+1, i+2, \dotsc, n \}$ with probability
 \begin{align*}
     \dfrac{\tau_{i, i'}(m)}{\sum_{i'': (i, i'') \in \A} \tau_{i, i''}(m)} \geq \dfrac{\tau_\text{min}(m)}{n \cdot \tau_\text{max}}.  
 \end{align*}
 As all the nodes reachable from $i$ to $t$ have already been $\varepsilon$-processed, the sum of the pheromone values of all the incorrect arcs $(i', i'')$ is at most $n \cdot \varepsilon$.
 Therefore, ant $a_i$ chooses the next arc from node $i'$ correctly with probability of at least $1 - n \cdot \varepsilon$ as the sum of the pheromone values of all the outgoing arcs of any node in $\mathcal{G}$ is at least $1$ according to Observation~\ref{lem:sumAtLeast1}.
 Thus, the probability that ant $a_i$ sees its optimal solution in iteration $m$, given that all the nodes reachable from $i$ to $t$ have already been $\varepsilon$-processed, is at least
 \begin{align*}
      \dfrac{\tau_\text{min}(m)}{n \cdot \tau_\text{max}} \cdot (1 - n \cdot \varepsilon)^{n-1}.
 \end{align*}
 As we have $\varepsilon = 1/n^2$, we get $(1 - n \cdot \varepsilon)^{n-1} \geq 1/e$.
\end{proof}

To proceed, we first need the following technical lemmas.
Let $\li(x) = \int_2^x \frac{1}{\ln t}\dd t$ denote the offset logarithmic integral function \cite[Chapter 5]{abramowitz+stegun}.

\begin{restatable}{lemma}{nastysum}
\label{lem:nastySum}
    For $0 < a \leq 1$ and $k \geq 3$, it holds that
    \[
        \sum_{m=1}^\infty e^{-a \li(m + k)}
            = e^{-a\li(k)} \cdot O\left(\frac{1}{a} \ln \left(\frac{1}{a}\li(k)\right) \right) \quad \text{as $a \to 0$ and $k \to \infty$}.
    \]
\end{restatable}

\begin{restatable}{lemma}{simplerecursion}\label{lem:recursion}
    Let $M_k$ be a sequence defined recursively for $k \geq 1$ as
    \[
        M_k = M_{k-1} + a \cdot \ln{M_{k-1}} + b
    \]
    with the initial value $M_1 = b$
    for $b \geq a > 0$. Then,
    \[
        M_k = O(b \cdot k + a \cdot k \cdot \ln{(a \cdot k)}).
    \]
\end{restatable}

To continue our analysis, we define the following random variables. 
Let $T_k$ denote the first cycle in which all nodes $i$ with $\ell(i) - \ell(t) \leq k$ are $\varepsilon$-processed.
Moreover, let $S_k$ denote the number of cycles \emph{after $T_{k-1}$}
needed for an ant $a_i$ with $\ell(i) - \ell(t) = k$ to see its optimal solution for the first time as a function of $T_{k-1}$.
Furthermore, let $\Delta_k$ denote the number of cycles required for a node $i$ with $\ell(i) - \ell(t) = k$ to become $\varepsilon$-processed after ant $a_i$ has seen its optimal solution for the first time. 
According to Lemma~\ref{lem:getProcessed1}, we have $\Delta_k = O\left(\frac{1}{\rho} \ln \dfrac{\tau_\text{max}}{\varepsilon}\right)$.

\begin{lemma}
\label{lem:firstTimeSeeOpt}
    The expected number of cycles until ant $a_i$ sees its optimal solution for the first time with $\ell(i) - \ell(t) = k$, given that all the nodes reachable from $i$ to $t$ have already been $\varepsilon$-processed, is $O\left(\dfrac{n \cdot \tau_\text{max}}{c_n} \cdot \ln{\dfrac{n \cdot \tau_\text{max} \cdot T_{k-1}}{c_n}}\right)$ when $\tau_\text{max} \geq 1$ and $\tau_\text{min}(m) = \frac{c_n}{\ln (m+1)}$, for $0 < c_n \leq \frac{1}{n^2}$.
\end{lemma}

\begin{proof}
     According to Lemma~\ref{lem:probOfSeeingOpt} We have
    \begin{align*}
            \exP[S_k \;\mid\; T_{k-1}] = \sum_{m=1}^\infty \probP (S_k \geq m \;\mid\; T_{k-1})
            \leq \sum_{m=1}^\infty \prod_{s=1}^{m-1} \left(1 - \dfrac{c \cdot \tau_\text{min}(T_{k-1} + s)}{n \cdot \tau_\text{max}}\right).
    \end{align*}
    As $\tau_\text{min}(m) = \frac{c_n}{\ln (m+1)}$, we get 
    \begin{align*}
            \exP[S_k \;\mid\; T_{k-1}] &\phantom{}\leq \sum_{m=1}^\infty \prod_{s=1}^{m-1} \left(1 - \dfrac{c \cdot c_n}{n \cdot \tau_\text{max} \cdot \ln(T_{k-1} + s + 1)}\right) \\
            &\leq \sum_{m=1}^\infty \exp \left(-\dfrac{c \cdot c_n}{n \cdot \tau_\text{max}} \sum_{s=1}^{m-1} \dfrac{1}{\ln(T_{k-1} + s + 1)} \right) \\
            &\leq \sum_{m=1}^\infty \exp \left(-\dfrac{c \cdot c_n}{n \cdot \tau_\text{max}} 
            \int_{s=T_{k-1}+2}^{m + T_{k-1} + 2} \dfrac{1}{\ln s} \dd s\right) \\
            &= \sum_{m=1}^\infty \exp \left(-\dfrac{c \cdot c_n}{n \cdot \tau_\text{max}} 
            \left(
                \li(m + T_{k-1} + 2) - \li(T_{k-1} + 2) 
            \right)\right) \\
            &\leq \exp\left(-\frac{c\cdot c_n}{n \tau_\text{max}}\cdot \li(T_{k-1}+2)\right)\sum_{m=1}^\infty \exp \left(-\dfrac{c \cdot c_n}{n \tau_\text{max}} 
                \cdot \li(m + T_{k-1} + 2)
            \right).
    \end{align*}
    %
    To invoke Lemma~\ref{lem:nastySum}, let $a = \dfrac{c \cdot c_n}{n \cdot \tau_\text{max}}$. 
    Since $\li(x) = O\left(\frac{x}{\ln x}\right)$ and $\ln\left(\frac{x}{\ln x}\right) \leq \ln x$ for $x \geq e$, we get, 
    \[
            \exP[S_k \;\mid\; T_{k-1}] = O\left(\dfrac{n \cdot \tau_\text{max}}{c_n} \cdot \ln{\dfrac{n \cdot \tau_\text{max} \cdot T_{k-1}}{c_n}}\right). \qedhere
    \]
\end{proof}

Finally, we provide an upper bound on the expected number of cycles to $\varepsilon$-process all the nodes in $\mathcal{G}$.

\nantrunningtime*
\begin{proof}
    Let $T$ be a random variable that indicates the total number of cycles in which all the nodes in $\mathcal{G}$ are $\varepsilon$-processed. We describe a recurrence relation that allows us to compute $\exP[T]$.
    This recurrence implicitly assumes that ant $a_i$ at node $i$ sees its shortest path
    before processing its node; hence, $\exP[T]$ also yields an upper bound to the number of
    cycles until all ants have seen their shortest paths.
    
    According to Lemma~\ref{lem:getProcessed1}, and since $\ell(1) - \ell(n) \leq n-1$, we have
    \begin{align*}
        \exP[T] \leq \exP[T_{n-1}]
        \leq \exP[T_{n-2}] + \exP[S_{n-1}] + \Delta_{n-1}
        = \exP[T_{n-2}] + \exP[S_{n-1}] + O\left(\frac{1}{\rho} \ln \dfrac{\tau_\text{max}}{\varepsilon}\right),
    \end{align*}
    where the second inequality arises as follows. Consider a node $i$ with $\ell(i) - \ell(t) = k$.
    For this node to be $\varepsilon$-processed, it suffices for $a_i$ to first see its shortest path and then wait
    for $\Delta_k$ cycles. After $T_{k-1}$, the ant $a_i$ is guaranteed to see its shortest path in $S_k$
    more cycles. 
    
    To compute $\exP[S_{n-1}]$, we use \Cref{lem:firstTimeSeeOpt} to find
    \[
        \exP[S_{n-1} \;\mid\; T_{n-2}] = O\left(
            \frac{n\cdot \tau_\text{max}}{c_n} \cdot \ln\left(\frac{n\cdot \tau_\text{max}}{c_n} \cdot T_{n-2}\right).
        \right).
    \]
    On using total expectation, we obtain the desired expectation; however, this requires us to 
    compute $\exP[\ln T_{n-2}]$. We therefore use Jensen's inequality, exploiting the concavity of the
    logarithm to obtain
    \begin{align*}
        \exP[S_{n-1}] = O\left(\dfrac{n \cdot \tau_\text{max}}{c_n} \cdot \ln{\dfrac{n \cdot \tau_\text{max}}{c_n}} + \dfrac{n \cdot \tau_\text{max}}{c_n} \cdot \ln \exP[T_{n-2}]\right).
    \end{align*}
    Therefore, we get
    \begin{align*}
        \exP[T_{n-1}]
        \leq \exP[T_{n-2}] + O\left(\dfrac{n \tau_\text{max}}{c_n} \cdot \ln \exP[T_{n-2}]\right) + O\left(\dfrac{n \tau_\text{max}}{c_n} \cdot \ln{\dfrac{n \tau_\text{max}}{c_n}} + \frac{1}{\rho} \ln \dfrac{\tau_\text{max}}{\varepsilon}\right).
    \end{align*}
    According to Lemma~\ref{lem:getProcessed1} and \ref{lem:firstTimeSeeOpt}, we have $\exP[T_1] = O\left(\dfrac{n \cdot \tau_\text{max}}{c_n} \cdot \ln{\dfrac{n \cdot \tau_\text{max}}{c_n}} + \dfrac{1}{\rho} \ln \dfrac{\tau_\text{max}}{\varepsilon}\right)$.
    By solving the recurrence relation for $\exP[T_{n-1}]$ based on~\Cref{lem:recursion}, we get
    \[
        \exP[T] = O\left(
            \frac{n^2 \cdot \tau_\text{max}}{c_n} \cdot \ln\left(\frac{n\cdot \tau_\text{max}}{c_n}\right)
                + \frac{n}{\rho} \ln\frac{\tau_\text{max}}{\varepsilon}
        \right).
    \]
    On simplifying the terms inside the $O(\cdot)$ and replacing $\varepsilon$ by $\frac{1}{n^2}$, the result follows.
\end{proof}

Furthermore, as any shortest path in $\mathcal{G}$ has at most $n$ nodes, the number of iterations in each cycle is upper-bounded by $O(n^2)$.

\begin{corollary}
\label{cor:totaliterations}
    Let $\mathcal{G} = (\mathcal{V}, \mathcal{A})$ be an instance of the SDSP on which we run $n$-ANT/tdlb with 
    $\tau_\text{min}(m) = c_n /\ln(m + 1)$. The expected number of iterations until all ants have found their shortest paths from their respective starting
    nodes in $\mathcal{G}$ is 
    $O\left(\dfrac{n^4 \cdot \tau_\text{max}}{c_n} \cdot \ln{\left(\dfrac{n\cdot \tau_\text{max}}{c_n} \right)} + \dfrac{n^3}{\rho} \ln(n\cdot\tau_\text{max})\right)$, for $\tau_\text{max} \geq 1$ and $0 < c_n \leq \frac{1}{n^2}$.
\end{corollary}

\Cref{thm:totalCycles} also yields the following, using the fact that the time
until all nodes are $\varepsilon$-processed is finite
with probability one (this follows from \Cref{thm:totalCycles} and Markov's inequality) and the properties
of the pheromone update. To be more precise, once all nodes are $\varepsilon$-processed,
then all ants have found some shortest path to the target node. Hence, the pheromone
on arcs not on these shortest paths will decrease geometrically to $\tau_{\min}$, which
itself decreases to zero. Meanwhile, the pheromone on the remaining arcs
is increased by $\rho$ in each subsequent cycle, eventually reaching a value of $\tau_{\max}$.

\nantlimitinggraph*

\section{Discussion}

In this paper, we have analyzed three variants of Ant Colony Optimization: GBAS/tdev, $n$-ANT/tdev, and $n$-ANT/tdlb.

\subparagraph{GBAS/tdev.} This ACO variant was first proposed and analyzed 
by Gutjahr~\cite{gutjahr2002aco}. Gutjahr left as an open question the running time of the algorithm. In this study, we obtained bounds on the running time that are generic, and hold for any combinatorial optimization problem. In addition, our analysis
shows that an evaporation rate of $\rho(m) = c/m$ for some constant $c$ suffices
to guarantee convergence. In contrast,
Gutjahr's analysis required an evaporation rate $\rho(m) \leq 1 - \ln(m) / \ln(m+1) = O\left(\frac{1}{m\ln m}\right)$.
Thus, we have a small improvement in the pheromone evaporation functions that yield
convergence.

Since the running time bound of \Cref{thm:gbas running time} is generic, it yields
rather large running times for specific problems. We expect that the running time can
be improved for specific problems. 

\subparagraph{{\boldmath$n$}-ANT/tdev.} \Cref{thm: nant running time tdev} tells
us that the running time of ACO algorithms with a time-dependent evaporation rate
may suffer compared to the simpler MMAS approach. Our construction
works for a wide range of evaporation rate functions, which includes the function
$\rho(m) = c/m$ that yields convergence for GBAS/tdev. To our knowledge, this is the first
time that the running time of the `tdev' modification has been analyzed rigorously.

\subparagraph{{\boldmath$n$}-ANT/tdlb.} \Cref{thm:totalCycles}
shows that the expected running time
of $n$-ANT/tdlb is polynomial on the single-destination shortest path problem (SDSP). Compared
to the original $n$-ANT algorithm, $n$-ANT/tdlb is not only guaranteed to find the optimal solution,
but also guarantees that the probability that any ant walks a non-shortest path approaches zero (\Cref{cor:limiting graph}).

The running time of $n$-ANT/tdlb according to \Cref{thm:totalCycles} is
somewhat worse than the running time of $n$-ANT obtained by 
Attiratanasunthron and Fakcharoenphol~\cite{Attiratanasunthron2008running}. However, we expect that this is more
an effect of the analysis than something one would witness in practice. After all,
for any polynomial number of iterations, we only decrease the pheromone lower bound
by a factor of $\ln n$ compared to $n$-ANT. Heuristically, this should only
make an expected $O(\ln n)$ factor difference in the running time.

While our results demonstrate that $n$-ANT/tdlb is efficient on the SDSP, the prime motivation
for ACO is to solve $\mathsf{NP}$-hard optimization problems, whereas the SDSP is polynomial-time
solvable. Thus, analyzing the `tdlb'/'tdev' modifications on harder optimization problems
may be of interest. In particular, can the analyses
of MMAS on the Travelling Salesperson Problem~\cite{Kötzing2012,zhou2009} be extended
to these time-dependent variants?

As a final remark, we note that the $n$-ANT algorithm was improved
by Sudholt and Thyssen \cite{sudholt2012running} to yield a better running time on
the SDSP. As the analysis becomes somewhat more involved, we did not analyze
their variant: our goal was not to determine the best ACO algorithm, but rather to show
that the tdlb method can yield polynomial running time algorithms. We thus leave open
the problem of extending Sudhold and Thyssen's algorithm using the tdlb method, and
to determine how this affects the running time.

\bibliography{refs}

\section{Technical Lemmas}

In this appendix, we prove the technical lemmas required in \Cref{sec:nant}.

\hardrecursion*

\begin{proof}
    Let $f : \mathbb{R} \to \mathbb{R}$ be a function satisfying
    $f(1) = M_1$ and
    \[
        f'(x) = c' \cdot f(x)^b,
    \]
    for some constant $c' < c$ to be determined later. First, observe that this
    differential equation is uniquely solved by
    \[
        f(x) = \left[
        M_1^{1-b} + c'(1-b)(x-1)
        \right]^{\frac{1}{1-b}}.
    \]
    For $x \geq 2$, we first write $f(x) = f(x-1) + \int_{x-1}^x f'(s)\dd s$ and use
    the differential equation above to find
    \begin{align*}
        f(x) &= f(x-1) + c'\int_{x-1}^x f(s)^b \dd x  \\
             &= f(x-1) + c'\int_{x-1}^x f(s-1)^b \left(\frac{f(s)}{f(s-1)}\right)^b \dd x
             \\
             &\leq f(x-1) + c'\cdot \max_{s \in [x-1, x]}
                \left(\frac{f(s)}{f(s-1)}\right)^b \int_{x-1}^x f(s-1)^b \dd x \\
             &\leq f(x-1) + c'\cdot 
             \max_{x \geq 2} \max_{s \in [x-1, x]} \left(\frac{f(s)}{f(s-1)}\right)^b 
             f(x-1)^b,
    \end{align*}
    since $f$ is monotone increasing and positive for for $s \geq 0$.

    We can compute the maximum above exactly:
    \begin{align*}
        \max_{x \geq 2} \max_{s \in [x-1, x]} \left(\frac{f(s)}{f(s-1)}\right)^b 
            &= \max_{x\geq 2}\max_{s \in [x-1, x]} \left(
                \frac{M_1^{1-b} + c'(1-b)(s-1)}
                    {M_1^{1-b} + c'(1-b)(s-2)}
            \right)^{\frac{b}{1-b}} \\
            &= \max_{x \geq 2} \left(
                \frac{M_1^{1-b} + c'(1-b)(x-2)}
                    {M_1^{1-b} + c'(1-b)(x-3)}
            \right)^{\frac{b}{1-b}} \\
            &=
            \left(
                \frac{M_1^{1-b}}
                    {M_1^{1-b} - c'(1-b)}
            \right)^{\frac{b}{1-b}}  \\
            &= \left(\frac{c}{c - c'}\right)^{\frac{b}{1-b}}.
    \end{align*}

    We now fix $c'$ by setting $c = c' \cdot \left(\frac{c}{c-c'}\right)^{\frac{b}{1-b}}$. We obtain with some
    algebraic manipulation an equation relating $c$ and $c'$,
    \[
        (c-c')^b c^{1-b} = (c')^{1-b} \cdot c^{b}.
    \]
    Although not obvious at first sight, this equation defines a line in the $(c, c')$-plane. Let $
    c' = \gamma_b c$; then the above yields in turn an equation for $\gamma_b$:
    \[
        c^b \cdot (1 - \gamma_b)^b c^{1-b} = \gamma_b^{1-b} \cdot c^{1-b} c^b \iff \gamma_b + \gamma_b^{\frac{1-b}{b}}
            = 1.
    \]
    
    Let $g(x) = x + x^{\frac{1-b}{b}}$. Since $g$ is continuous and monotone increasing, and $g(0) = 0$
    and $g(1) = 2$, there exists by the intermediate value theorem a unique $\gamma_b \in (0, 1)$ such
    that $g(\gamma_b) = 1$.
    Thus, the above equation has a solution in $(0, 1)$ for every fixed $b \in (0, 1)$.

    Let $L_k = f(k)$ for $k \geq 1$. All together, the above yields that $L_k$ satisfies the recurrence
    inequality
    \[
        L_k \leq L_{k-1} + c \cdot (L_{k-1})^b
    \]
    with $L_1 = M_1$. Consequently, for all $k \geq 1$ we have $L_k \leq M_k$, thus completing the proof.
\end{proof}

\lowerboundbsmall*

\begin{proof}
    We have
    \[
    \left(x^{b} + a\right)^{\frac{1}{b}} - x  = x\left(1+\frac{a}{x^{b}} \right)^{\frac{1}{b}} - x
    \geq x\left(1 + \frac{1}{b}\left( \frac{a}{x^{b}} \right) \right) - x
    = \frac{a}{b} \cdot x^{1-b}. \qedhere
    \]
\end{proof}

\nastysum*

\begin{proof}
   We start by noting that the logarithmic integral function $\li(x)$ is monotone increasing for $x > 1$.
   Hence, we can bound the sum as
    \[
        \sum_{m=1}^\infty \exp\left(-a\li(m+k)\right) \leq 
            \int_0^\infty  \exp\left(-a \li(m + k)\right) \dd m
            = \int_{k}^\infty  \exp\left(-a\li(m)\right) \dd m.
    \]
    Now we perform a variable substitution, setting $x = \li(m)$.
    Since $\li$ is monotone increasing on $[k, \infty)$ it is also one-to-one; hence, we can use standard integral transformation rules,
    obtaining for the integral above
    \[
        \int_{\li(k)}^\infty \exp\left(-ax\right) \left|(\li^{-1})'(x)\right|\dd x,  
    \]
    By the inverse function theorem, 
    \[
        (\li^{-1})'(x) = \frac{1}{\li'(\li^{-1}(x))} = \ln \li^{-1}(x),
    \]
    since $\li'(x) = \frac{\dd}{\dd x} \int_0^x \frac{1}{\ln t}\dd t = \frac{1}{\ln x}$.
    Thus,
    \begin{align}\label{eq: integral involving inverse}
        \int_{\li(k)}^\infty \exp\left(-ax\right) \left|(\li^{-1})'(x)\right|\dd x
            =  \int_{\li(k)}^\infty \exp\left(-ax\right) 
                    \left|\ln \li^{-1}(x)\right|\dd x.
    \end{align}
    
    Next, we bound the factor involving $\li^{-1}$, which we can do without explicitly computing
    $\li^{-1}$. We aim to show that there exists a constant $c > 0$ such that
    \[
        \left|\ln \li^{-1}(x)\right| \leq c\ln x
    \]
    for $x \in \left[\li(k), \infty\right)$. By construction, we have 
    $\li^{-1}\left(\li\left(y\right)\right) = y$. Thus, we set $x = \li(y)$. Proving the above
    is then equivalent to showing that $\ln y \leq c \ln \li(y)$ for $y \geq k$.
    Since $\li(y) \geq \frac{y}{2\ln y}$ for all $y \geq 3$, it suffices to show that
    \[
        \ln y \leq c \cdot \ln \left(\frac{y}{2\ln y}\right)
            = c \cdot \left(\ln y - \ln \ln y - \ln 2\right).
    \]
    
    On $[k, \infty) \subseteq [3, \infty)$, it holds that $\ln y - \ln \ln y - \ln 2 > \frac{1}{4} \ln y$.
    Thus, taking $c = 4$ suffices.

    We thus replace the factor involving $f^{-1}$ in \Cref{eq: integral involving inverse}
    by $4\ln x $.
    We then perform another integral substitution, letting $z = a\left(x -\li(k)\right)$.
    We subsequently split the integral into two terms:
    \begin{multline*}
            \int_{\li(k)}^\infty e^{-ax} \ln x \dd x \\
                = 
            \frac{1}{a}e^{-a\li(k)}\left(
            \int_0^{a \li(k)}e^{-z} \ln
                \left( \frac{z}{a} + \li(k) \right) \dd z
                +
            \int_{a\li(k)}^\infty e^{-z} \ln
                \left( \frac{z}{a} + \li(k) \right) \dd z
            \right).
    \end{multline*}
    We treat the terms separately. First, observe that  
    \begin{align*}
            \int_0^{a\li(k)} e^{-z} \ln
                \left( \frac{z}{a} + \li(k) \right) \dd z
                    \leq \ln\left(2\li(k)\right) \cdot \int_0^\infty e^{-z} \dd z
                    = O\left(\ln\left(\li(k)\right)\right),
    \end{align*}
    yielding a bound for the first integral. In the second integral, through the entire integration domain
    it holds that $\li(k) \leq z/a$. Using this fact and the fact that $\li(k) > 1$
    for $k \geq 3$ allows us to bound the integral as follows:
    \begin{align*}
            \int_{a\li(k)}^\infty e^{-z} \ln
                \left( \frac{z}{a} + \li(k) \right) \dd z
            &\leq
                \int_{a\li(k)}^\infty e^{-z} \ln \left(\frac{2z}{a}\right)\dd z \\
            &\leq 
                \int_{a}^\infty e^{-z}\ln \left(\frac{2z}{a}\right)\dd z \\
            &= 
                O\left(\ln\frac{1}{a}\right) + \int_{a}^\infty e^{-z}\ln (2z)\dd z\\
            &= 
            O\left(\ln\frac{1}{a}\right).
    \end{align*}
    Finally, putting the two terms together completes the proof.
\end{proof}

\simplerecursion*

\begin{proof}
    Let $D_k = M_k - M_{k-1} = a \cdot \ln{M_{k-1}} + b$. Then,
    \begin{align}
    \label{eq:recursionOne}
        M_k &= M_1 + \sum_{j=2}^{k}D_j
        = b + \sum_{j=2}^{k} (a \ln{M_{j-1}} + b)
       = b \cdot k + a \cdot \sum_{j=2}^{k} \ln{M_{j-1}} \nonumber \\
        &\leq b \cdot k + a \cdot k \cdot \ln{M_{k}}.
    \end{align}
    This inequality is solved by
    \[
        -a\cdot k \cdot W_{0}\left(-\frac{e^{-b/a}}{a \cdot k}\right) \leq M_k 
            \leq -a\cdot k \cdot W_{-1}\left(-\frac{e^{-b/a}}{a\cdot k}\right),
    \]
    where $W_0$ and $W_{-1}$ denote the upper and lower branches of the Lambert W 
    function~\cite[Section 4.13]{NIST:DLMF}. 
    Using a bound by Chatzigeorgiou~\cite[Theorem 1]{chatzigeorgiou2013}:
    \[
        W_{-1}\left(-e^{-u-1}\right) > -1 - \sqrt{2u} - u \quad \text{for $u > 0$},
    \]
    we find
    \[
        M_k \leq a \cdot k \cdot O\left(1 + \frac{b}{a} + \ln(a\cdot k)\right)
            = O\left(b \cdot k + a \cdot k \cdot \ln (a \cdot k)\right). \qedhere
    \]
\end{proof}

\appendix

\end{document}